\documentclass[
  journal=pasa,
  manuscript=research-paper,
  year=2025,
  volume=YY
]{cup-journal}

\usepackage{tikz}
\usepackage{xcolor} 
\usepackage{amssymb} 
\usepackage[pdfpagelabels=false]{hyperref}
\hypersetup{
    colorlinks=true,
    linkcolor=blue,
    citecolor=blue,
    filecolor=blue,
    urlcolor=blue
}

\newcommand{\bluesquare}{\tikz \filldraw[fill={rgb,255:red,3; green,2; blue,255}] (0,0) rectangle (0.25,0.25); }

\newcommand{\shadeduptriangle}{\tikz \filldraw[black] (0,0) -- (0.3,0) -- (0.15,0.26) -- cycle;}
\newcommand{\shadeddowntriangle}{\tikz \filldraw[black] (0,0.3) -- (0.3,0.3) -- (0.15,0) -- cycle;}

\newcommand{\fewcorot}{\hbox{$f_{\rm EWcorot}$}}
\newcommand{\slope}{\hbox{${d{(\fewcorot)}}/{d{(\rm eV)}}$}}

\newcommand{\dfcorot}{\hbox{$\Delta \fewcorot (\rm{\MgII} - \rm {\OVI})$}}

\newcommand{\magiicat}{\hbox{{\rm MAG}{\sc ii}CAT}}
\newcommand{\Ha}{H$\alpha$}

\newcommand{\PVdblt}{{\rm P}\kern 0.1em{\sc v}~$\lambda\lambda 1117, 1128$}
\newcommand{\CaIIdblt}{{\rm Ca}\kern 0.1em{\sc ii}~$\lambda\lambda 3934, 3969$}
\newcommand{\AlIIIdblt}{{\rm Al}\kern 0.1em{\sc iv}~$\lambda\lambda 1855, 1863$}
\newcommand{\CIVdblt}{{\rm C}\kern 0.1em{\sc iv}~$\lambda\lambda 1548, 1550$}
\newcommand{\MgIIdblt}{{\rm Mg}\kern 0.1em{\sc ii}~$\lambda\lambda 2796, 2803$}
\newcommand{\NVdblt}{{\rm N}\kern 0.1em{\sc v}~$\lambda\lambda 1238, 1242$}  
\newcommand{\SVIdblt}{{\rm S}\kern 0.1em{\sc vi}~$\lambda\lambda 933, 944$} 
\newcommand{\OVIdblt}{{\rm O}\kern 0.1em{\sc vi}~$\lambda\lambda 1031, 1037$} 
\newcommand{\SiIIdblt}{{\rm Si}\kern 0.1em{\sc ii}~$\lambda\lambda 1190, 1193$} 
\newcommand{\SiIVdblt}{{\rm Si}\kern 0.1em{\sc iv}~$\lambda\lambda 1393, 1402$} 
\newcommand{\PV}{\hbox{{\rm P}\kern 0.1em{\sc v}}}
\newcommand{\AlI}{\hbox{{\rm Al}\kern 0.1em{\sc i}}}
\newcommand{\AlII}{\hbox{{\rm Al}\kern 0.1em{\sc ii}}}
\newcommand{\AlIII}{{\hbox{\rm Al}\kern 0.1em{\sc iii}}}
\newcommand{\CaII}{\hbox{{\rm Ca}\kern 0.1em{\sc ii}}}
\newcommand{\CII}{\hbox{{\rm C}\kern 0.1em{\sc ii}}}
\newcommand{\CIIe}{\hbox{{\rm C$^{\ast}$}\kern 0.1em{\sc ii}}}
\newcommand{\CIII}{\hbox{{\rm C}\kern 0.1em{\sc iii}}}
\newcommand{\CIV}{\hbox{{\rm C}\kern 0.1em{\sc iv}}}
\newcommand{\CV}{\hbox{{\rm C}\kern 0.1em{\sc v}}}
\newcommand{\HI}{\hbox{{\rm H}\kern 0.1em{\sc i}}}
\newcommand{\HII}{\hbox{{\rm H}\kern 0.1em{\sc ii}}}
\newcommand{\Lya}{\hbox{{\rm Ly}\kern 0.1em$\alpha$}}
\newcommand{\Lyb}{\hbox{{\rm Ly}\kern 0.1em$\beta$}}
\newcommand{\Lyg}{\hbox{{\rm Ly}\kern 0.1em$\gamma$}}
\newcommand{\Lyd}{\hbox{{\rm Ly}\kern 0.1em$\delta$}}
\newcommand{\Lye}{\hbox{{\rm Ly}\kern 0.1em$\epsilon$}}
\newcommand{\Lyphi}{\hbox{{\rm Ly}\kern 0.1em$\phi$}}
\newcommand{\Lyfive}{\hbox{{\rm Ly}\kern 0.1em$5$}}
\newcommand{\Lysix}{\hbox{{\rm Ly}\kern 0.1em$6$}}
\newcommand{\Lyseven}{\hbox{{\rm Ly}\kern 0.1em$7$}}
\newcommand{\Lyeight}{\hbox{{\rm Ly}\kern 0.1em$8$}}
\newcommand{\Lynine}{\hbox{{\rm Ly}\kern 0.1em$9$}}
\newcommand{\Lyten}{\hbox{{\rm Ly}\kern 0.1em$10$}}
\newcommand{\Lyeleven}{\hbox{{\rm Ly}\kern 0.1em$11$}}
\newcommand{\HeI}{\hbox{{\rm He}\kern 0.1em{\sc i}}}
\newcommand{\HeII}{\hbox{{\rm He}\kern 0.1em{\sc ii}}}
\newcommand{\FeI}{\hbox{{\rm Fe}\kern 0.1em{\sc i}}}
\newcommand{\FeII}{\hbox{{\rm Fe}\kern 0.1em{\sc ii}}}
\newcommand{\FeIII}{\hbox{{\rm Fe}\kern 0.1em{\sc iii}}}
\newcommand{\MnII}{\hbox{{\rm Mn}\kern 0.1em{\sc ii}}}
\newcommand{\MgI}{\hbox{{\rm Mg}\kern 0.1em{\sc i}}}
\newcommand{\MgIb}{\hbox{{\rm Mg}\kern 0.1em{\sc i}}\kern 0.05em{\rm b}}
\newcommand{\MgII}{\hbox{{\rm Mg}\kern 0.1em{\sc ii}}}
\newcommand{\MgIII}{\hbox{{\rm Mg}\kern 0.1em{\sc iii}}}
\newcommand{\NI}{\hbox{{\rm N}\kern 0.1em{\sc i}}}
\newcommand{\NII}{\hbox{{\rm N}\kern 0.1em{\sc ii}}}
\newcommand{\NIII}{\hbox{{\rm N}\kern 0.1em{\sc iii}}}
\newcommand{\NV}{\hbox{{\rm N}\kern 0.1em{\sc v}}}
\newcommand{\OVI}{\hbox{{\rm O}\kern 0.1em{\sc vi}}}
\newcommand{\OI}{\hbox{{\rm O}\kern 0.1em{\sc i}}}
\newcommand{\OII}{\hbox{[{\rm O}\kern 0.1em{\sc ii}]}}
\newcommand{\OIII}{\hbox{[{\rm O}\kern 0.1em{\sc iii}]}}
\newcommand{\OIV}{\hbox{{\rm O}\kern 0.1em{\sc iv}]}}
\newcommand{\SI}{{\rm S}\kern 0.1em{\sc i}}
\newcommand{\SIV}{{\rm S}\kern 0.1em{\sc iv}}
\newcommand{\SVI}{{\rm S}\kern 0.1em{\sc vi}}
\newcommand{\SiI}{\hbox{{\rm Si}\kern 0.1em{\sc i}}}
\newcommand{\SiII}{\hbox{{\rm Si}\kern 0.1em{\sc ii}}}
\newcommand{\SiIII}{\hbox{{\rm Si}\kern 0.1em{\sc iii}}}
\newcommand{\SiIV}{\hbox{{\rm Si}\kern 0.1em{\sc iv}}}
\newcommand{\SII}{\hbox{{\rm S}\kern 0.1em{\sc ii}}}
\newcommand{\SIII}{\hbox{{\rm S}\kern 0.1em{\sc iii}}}
\newcommand{\NaI}{\hbox{{\rm Na}\kern 0.1em{\sc i}}}
\newcommand{\NaID}{\hbox{{\rm Na}\kern 0.1em{\sc i}}\kern 0.05em{\rm D}}
\newcommand{\TiII}{\hbox{{\rm Ti}\kern 0.1em{\sc ii}}}
\newcommand{\kms}{\hbox{~km~s$^{-1}$}}

\newcommand{\Rv}{\hbox{$R_{\rm vir}$}}

\usepackage{amsmath}
\usepackage[nopatch]{microtype}
\usepackage{booktabs}
\usepackage{graphicx}
\usepackage{subcaption}
\usepackage{natbib}
\bibpunct{(}{)}{;}{a}{}{;} 
    
\title{COS-EDGES: Co-rotation and Kinematic Stratification of the Multi-Phase CGM Around Edge-On Galaxies}

\author{Glenn G. Kacprzak}
\affiliation{Centre for Astrophysics and Supercomputing, Swinburne University of Technology, Hawthorn, Victoria 3122, Australia}
\author{Benjamin D. Oppenheimer}
\affiliation{University of Colorado, Center for Astrophysics and Space Astronomy, 389 UCB, Boulder, CO 80309, USA}
\author{Nikole M. Nielsen}
\affiliation{Homer L. Dodge Department of Physics and Astronomy, The University of Oklahoma, 440 W. Brooks St., Norman, OK 73019, USA}
\alsoaffiliation{Centre for Astrophysics and Supercomputing, Swinburne University of Technology, Hawthorn, Victoria 3122, Australia}
\author{Antonia Fern\'{a}ndez-Figueroa}
\affiliation{Centre for Astrophysics and Supercomputing, Swinburne University of Technology, Hawthorn, Victoria 3122, Australia}
\author{Michael T. Murphy}
\affiliation{Centre for Astrophysics and Supercomputing, Swinburne University of Technology, Hawthorn, Victoria 3122, Australia}
\author{Rebecca J. Allen}
\affiliation{Centre for Astrophysics and Supercomputing, Swinburne University of Technology, Hawthorn, Victoria 3122, Australia}
\author{Tania M. Barone}
\affiliation{Centre for Astrophysics and Supercomputing, Swinburne University of Technology, Hawthorn, Victoria 3122, Australia}
\author{Sameer}
\affiliation{Department of Physics and Astronomy, University of Notre Dame, Notre Dame, IN 46556, USA}
\author{Christopher W. Churchill}
\affiliation{Department of Astronomy, New Mexico State University, 1320 Frenger Mall, Las Cruces, NM 88003-8001, USA}
\author{Joseph N. Burchett}
\affiliation{Department of Astronomy, New Mexico State University, 1320 Frenger Mall, Las Cruces, NM 88003-8001, USA}
\author{Kaustubh R. Gupta}
\affiliation{Centre for Astrophysics and Supercomputing, Swinburne University of Technology, Hawthorn, Victoria 3122, Australia}
\author{Jane C. Charlton}
\affiliation{Department of Astronomy and Astrophysics, The Pennsylvania State University, State College, PA 16801, USA}
\author{Caleb B. Platukis}
\affiliation{Department of Astronomy and Astrophysics, The Pennsylvania State University, State College, PA 16801, USA}
\email[Glenn Kacprzak]{gkacprzak@swin.edu.au}


\begin{document}

\begin{abstract}

We present the first results from the COS-EDGES survey, targeting the kinematic connection between the interstellar medium and multi-phase circumgalactic medium (CGM) in nine isolated, near-edge-on galaxies at $z\sim0.2$, each probed along its major axis by a background quasar at impact parameters of $D=13-38$~kpc. Using VLT/UVES and HST/COS quasar spectra, we analyse {\MgI}, {\MgII}, {\HI}, {\CII}, {\CIII}, and {\OVI} absorption relative to galaxy rotation curves from Keck/LRIS and Magellan/MagE spectra. We find that low ionisation absorption for 8/9 galaxies lies below the halo escape velocity, indicating bound inflow or recycling gas, while 6/9 galaxies have high ionisation gas reaching above the halo escape velocity, suggesting some unbound material. We find that at lower $D/R_{\rm vir}$ ($0.12\leq$$D/R_{\rm vir}$$\leq0.20$), over $80\%$ of absorption in all ions lies on the side of systemic velocity matching disk rotation, and the optical-depth–weighted median velocity ($v_{abs}$) is consistent with the peak rotation speed. At higher $D/R_{\rm vir}$ ($0.21\leq$$D/R_{\rm vir}$$\leq0.31$), the kinematics diverge by ionisation state: For the low ionisation gas, the amount of co-rotating absorption remains above 80\%, yet $v_{abs}$ drops to roughly 60\% of the galaxy rotation speed. For the high ionisation gas ({\OVI}), only 60\% of the absorption is consistent with co-rotation and $v_{abs}$ drops to 20\% of the galaxy rotation speed. Furthermore, the velocity widths, corresponding to 50\% of the total optical depth ($\Delta v_{50}$) for low ionisation gas is up to 1.8 times larger in the inner halo than at larger radii, while for {\CIII} and {\OVI} $\Delta v_{50}$ remains unchanged with distance. The 90\% optical-depth width ($\Delta v_{90}$) shows a modest decline with radius for low ionisation gas but remains constant {\CIII} and {\OVI}. At high $D/\Rv$ both $\Delta v_{50}$ and $\Delta v_{90}$ increase with ionisation potential. These results suggest a radially dependent CGM kinematic structure: the inner halo hosts cool, dynamically broad gas tightly coupled to disk rotation, whereas beyond $\gtrsim 0.2 R_{\rm vir}$, particularly traced by {\OVI} and {\HI}, the CGM shows weaker rotational alignment and lower relative velocity dispersion.  Therefore, low-ionisation gas likely traces extended co-rotating gas, inflows and/or recycled accretion, while high-ionisation gas reflects a mixture of co-rotating, lagging, discrete collisionally ionised structures and volume-filling warm halo, indicating a complex kinematic stratification of the multi-phase CGM.

\end{abstract}


\section{Introduction} In 1986, a galaxy near a quasar sightline was determined to have a redshift matching {\MgII} absorption found in the quasar's spectrum \citep{bergeron86}, providing a key observational link between a galaxy and its circumgalactic medium (CGM) \citep[also see][]{haschick75, boksenberg78}. Since Jacqueline Bergeron's finding, significant efforts have been made to understand the connection between the CGM and its host galaxy and its role in influencing galaxy evolution.


The kinematics of the CGM (velocity centroids, dispersions, and systemic redshifts) can be directly measured from absorption profiles and provide powerful insights into the origins of this gas. Early studies showed that {\MgII} absorption systems typically exhibited velocity structures where clouds clustered together, occasionally with higher-velocity clouds offset from the main groupings. These coherent motions were proposed to arise from rotating halos \citep{weisheit78,lanzetta92} or from a combination of extended disk rotation and infalling accretion \citep{charlton98}. Additionally, \citet{prochaska97} determined that the metal-line kinematics of Damped {\Lya} systems (DLAs) strongly favoured a cold, disk-like origin. These early results highlighted the need to understand how absorption kinematics relate to their host galaxies, while further determining if the gas is co-rotating or infalling.


Further insights into the connections between the CGM and their galaxies came with the morphological modelling of {\MgII} host galaxy properties using high resolution images \citep{kacprzak07}. Since then, studies have shown that the spatial distribution of the CGM of intermediate redshift galaxies is bimodal; gas is more frequently detected along the major and minor axes of galaxies for ions such as {\MgII} \citep{bordoloi11,bouche12,kacprzak12,lan14,martin19,zabl19,lundgren21}, {\OVI} \citep{kacprzak15,beckett21} and {\HI} \citep{beckett21}. These results support the idea that gas flows out of galaxies along their minor axes, while gas accretes along the disk plane, which is the picture supported by simulations \citep{peroux20,defilippis21,pillepich21,hafen22,trapp22,yang24}. Thus, to fully understand the gas flow processes found in the CGM, both kinematics and the galaxy-quasar geometry should be examined.  

Simulations show that the low-ionisation CGM most likely co-rotates with the galaxy disk as it accretes onto galaxies \citep{stewart13,stewart17,ho19}. This is confirmed by {\rm Mg}\kern 0.1em{\sc ii} absorption found along the major axes of galaxies, which typically aligns with the direction of galaxy rotation out to 100~kpc \citep{steidel02,kacprzak10,kacprzak11kin,bouche13,bouche16,diamond-stanic16,ho17, rahmani18, martin19,lopez20}. Similar co-rotating signatures have been found for {\Lya} \citep{barcons95,french20,nateghi24GFI}, which is expected since {\HI} can trace the low ionisation phase. Furthermore, the kinematic consistency between the galaxy and {\Lya} is at its highest within the galaxy virial radius and decreases with increasing distance \citep{nateghi24GFI}. This kinematic consistency further decreases with decreasing {\HI} column density \citep{nateghi24GFI}. 

While {\HI} mostly traces the cool low ionisation gas phase, it is associated with warmer gas phases as well. Thus, it is important to address whether all ions are tracing the same kinematics or arising from different structures from the CGM. \citet{kacprzak19kine} reported that {\OVI}, found along the major axes of ten galaxies with impact parameters ranging between $35 \leq D \leq 276$~kpc ($0.14 \leq D/R_{\rm vir} \leq 1.3$), appears to be less kinematically connected to the rotation of the galaxy, with {\OVI} mostly spanning the galaxy systemic velocities. Using cosmological simulations, they showed that while {\OVI} did exhibit signatures of accretion along the major axis, they concluded that the observational inflow signatures were obscured by the broad distribution of the {\OVI} velocities across the halo. Other studies further showed that a significant fraction of {\OVI} detected in absorption systems associated with galaxies actually resides outside of the galaxy halo \citep{ho21,bromberg24}. Yet, signatures of {\OVI} co-rotation do exist, and when {\OVI} is consistent with a co-rotation model, these absorbers primarily reside along the galaxies’ major axis \citep{nateghi24GFII}. \citet{ho25}, in a sample of 18 edge-on galaxies probed along their major axes with impact parameters ranging between $20 \leq D \leq 231$~kpc ($0.12 \leq D/R_{\rm vir} \leq 1.19$), showed that although {\OVI} clouds at large radii can display complex kinematics, the majority of the warm–hot halo gas nevertheless co-rotates with the galaxy and seldom exhibits retrograde motion. This implies that even  {\OVI} retains a substantial fraction of the galaxy’s angular momentum. Thus, addressing how {\OVI} and other ions are kinematically connected to their galaxies will shed light on the physical processes within the CGM.

A recent study by  \citet{nateghi24GFII} inspected 27 galaxies ($0.09 < z < 0.5$)  associated with multi-phase absorption at impact parameters  ranging between $21 \leq D \leq 276$~kpc ($0.16 \leq D/R_{\rm vir} \leq 1.61$) and found that the amount of co-rotating gas decreases with increasing ionisation potential. This showed that the low-ionisation CGM is more kinematically coupled to the galaxy  compared to the high-ionisation CGM. They further reported that the fraction of 
co-rotating low and high ionisation potential species likely increases with increasing {\HI} column density, but this correlation is likely shallower for higher ionisation gas.
They further showed that the co-rotation of lower ionisation gas decreases with increasing distance away from the galaxy \citep[also consistent with][]{klimenko23} while {\OVI} was consistent with being flat. These studies have provided useful information regarding the kinematic connection between galaxies and their multi-phase CGM, yet their samples contain a range of galaxy properties, redshifts and orientations, which could result in probing a range of CGM phenomena and masking stronger correlations. In order to target specific CGM processes, such as gas accretion, more focused samples are required.

The aim of our program is to address how gas flows towards galaxies via an analysis of the galaxies and their multi-phase CGM. To do this, we have selected a sample of $z\sim0.2$, isolated, edge-on galaxies with the quasar located along their major axis at projected distances of $D=13-38$~kpc. In this first paper, we compare the galaxies’ rotation velocities to the CGM absorption kinematics across different ions, assessing the presence or absence of co-rotation as a function of gas phase and distance. The paper is organised as follows: In Section~\ref{sec:method} we present galaxy properties and the data used in our analysis.  In Section~\ref{sec:results} we present a kinematic analysis of the ISM and the CGM. We discuss our results in Section~\ref{sec:discussion} and present our concluding remarks in Section~\ref{sec:conclusion}. The
Appendices include the individual galaxy-CGM kinematic comparisons for reference. Throughout this paper, we adopt an ${\rm H}_{\rm 0}=70$~\kms~Mpc$^{-1}$, $\Omega_{\rm M}=0.3$, $\Omega_{\Lambda}=0.7$ cosmology.

\begin{figure*}[hbt!]
\centering
\begin{subfigure}{0.49\linewidth}
    \centering
    \includegraphics[width=\linewidth]{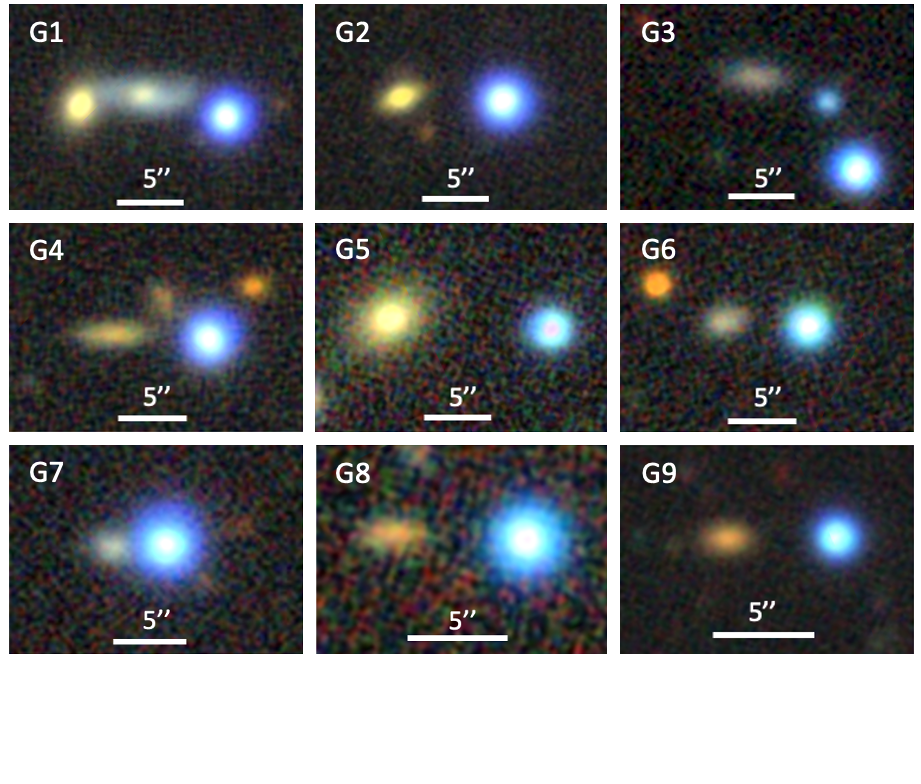}
\end{subfigure}
\hfill
\begin{subfigure}{0.49\linewidth}
    \includegraphics[width=\linewidth]{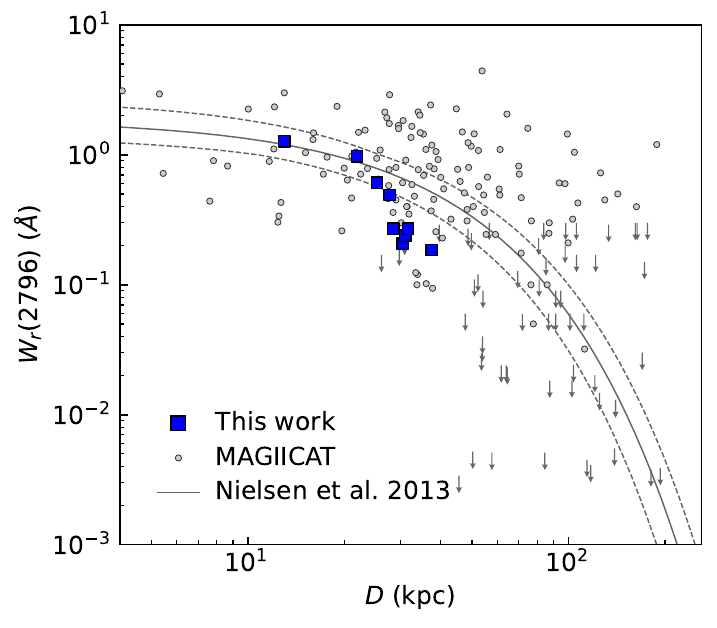}
 \end{subfigure}
\caption{(Left) DECaLS $grz$ composite images of each quasar field, in which the quasar appears as a bright blue point source and the largest galaxy in the frame is the targeted foreground galaxy. The galaxies are moderately inclined and have their major axes pointed towards the quasar sightline. These galaxies were previously identified as being isolated \citep{huang21}. (Right) Impact parameter, $D$, versus the {\MgII} $\lambda$2796 rest-frame equivalent width, $W_r(2796)$. The blue squares are the nine galaxies examined here. The grey data is from {\magiicat} \citep{magiicat1}, where galaxies having detected {\MgII} absorption are presented as circles and those with upper limits on absorption are downward arrows. The solid curve is a log-linear maximum likelihood fit to the {\magiicat} data with 1$\sigma$ uncertainties shown as dotted lines, which were adopted from \citet{magiicat2}. 
 }
\label{fig:EWD}
\end{figure*}
\section{Data and Analysis}
\label{sec:method}
The following section presents the COS-EDGES survey and data obtained for this sample of nine galaxies. The galaxies were selected from \citet{chen10a} and \citet{huang21} and were chosen to be moderately inclined ($60\leq i \leq 90$ degrees) with the major axis closely aligned with the quasar sightline (1 $\leq \Phi \leq $ 35 degrees) with detected {\MgII} absorption.  The galaxies have a stellar mass range of $\log(M_{\star}/M_{\odot})=9.4-10.6$ and are within 38~kpc, or 0.31\Rv, of the background quasars.  We selected targets with low galaxy-quasar separations in order to examine how gas transitions from the CGM to the ISM. Images of the nine galaxies are presented in Figure~\ref{fig:EWD} and the galaxy properties are shown in Table~\ref{tab:gal_params}.

\subsection{VLT/UVES Quasar Spectra}
The {\MgII} absorption associated with these galaxies was already identified using low/moderate resolution spectra \citep{chen10a,huang21}. However, in order to examine the individual kinematic components of each absorption system, high resolution spectra are required. These data were obtained using UVES \citep{dekker-uves} which is a dual-arm, grating cross-dispersed high resolution optical echelle spectrograph mounted on the Nasmyth B focus of Unit Telescope 2 of the VLT. The program IDs for these data are 105.20FN.001, 105.20FN.002 and 108.22F4.002. The data were taken between 2020 and 2022 using a slit width of 1.2$''$ and using the Blue CCD1 setting in order to capture {\MgII}, {\MgI} and {\FeII} over the wavelength range of 3000-4000~{\AA}. The exposure times were based on quasar brightness in the U-band and range in exposure times between $1,434-11,200$~seconds. The UVES spectra were reduced using a combination of the standard ESO pipeline and the UVES Post–Pipeline Echelle Reduction \citep[\textsc{uves popler};][]{uvespopler,murphy2019}. The spectra were both vacuum and heliocentric corrected and have a final velocity dispersion of 2.5~{\kms}~pixel$^{-1}$.  Analysis of the {\MgII} and {\MgI} absorption profiles was performed using our own interactive software \citep[see][]{cv01,churchill20} for local continuum fitting, objective feature identification, and measuring absorption properties. 

\subsection{HST/COS Quasar Spectra}
The Hubble Space Telescope (HST) Cosmic Origins Spectrograph \citep[COS;][]{green12} was used to obtain the ultraviolet spectra of each background quasar (PID 17541). We optimised the G130M grating ($R\sim20,000$, (FWHM$\sim18$~{\kms}) central wavelength for each target, given the galaxy redshift, to maximise coverage of a range of ions ({\CII}, {\CIII}, {\NII}, {\NIII}, {\OI}, {\OVI}, {\SiII} and {\FeII}) and coverage of the Lyman series from {\Lyb} onwards, using the 1091, 1222 and 1291 settings. Data\footnote{The data described here may be obtained from \url{http://dx.doi.org/10.17909/bcdb-6c02}} were retrieved from the Barbara A. Mikulski Archive for Space Telescopes (MAST).  We binned the spectra by three pixels to increase the signal-to-noise ratio (S/N) and our analysis was performed on the binned spectra. Analysis of the absorption systems was performed using our own interactive software \citep[see][]{cv01} for local continuum fitting, objective feature identification, and measuring absorption properties.

\subsection{Galaxy Spectra}
We obtained spectra of the nine galaxies using either Magellan/MagE or Keck/LRIS.  

Galaxies G1, G8 and G9 were obtained using the MagE spectrograph \citep{marshall08} mounted on the Magellan I Baade Telescope (6.5 m) at the Las Campanas Observatory in Chile on 17 and 18 of August 2018. MagE covers the observed-frame optical ($\sim 3200-8500$~{\AA}) wavelengths. We used a 0.7$''$ slit with $1\times1$ binning, resulted in a spectral resolution of $R\sim5900$. The spatial resolution is $0.3''/$pixel. Exposure times ranged between 2400 and 4200 seconds. Data was reduced using the PypeIt reduction pipeline \citep{prochaska20pypit}, using standard stars and arc lamps taken during the night. The spectra were vacuum and heliocentric corrected. 

Spectra for the remaining six galaxies were obtained using the LRIS spectrograph \citep{oke95,mccarthy98} mounted on the Keck I Telescope (10 m) at the Keck Observatory in Hawaii on 20 January 2023. We used the B600/4000 grism ($R\sim 1300$) for LRIS Blue and R1200/7500 grating ($R\sim 4700$) for LRIS Red with a 1$''$ wide slit with $1\times1$ binning.  The spatial resolution is $0.135''/$pixel. The data were reduced using IRAF, with standard stars and arc lamps taken during the night.
The spectra were vacuum and heliocentric corrected. 

We extracted the galaxy rotation curves by employing a three-pixel-wide aperture size and shifted it at one-pixel intervals along the spatial direction to extract a sequence of spectra along the galaxy major axis \citep[see][]{vogt96,steidel02,kacprzak10}. Gaussian profiles were fitted to the galaxy emission lines using {\Ha}, or  {\OII} when {\Ha} was blended with skylines, to determine their wavelengths and line-of-sight velocity centroids within an accuracy of less than 6~{\kms}. These centroids were then utilized to determine the galaxy systemic redshifts and rotation curves. 

Star formation rates (SFRs) were computed using {\Ha} line fluxes, or {\OII} line fluxes when {\Ha} was blended with skylines, using the relations from \citet{kewley04}. We corrected the lines fluxes for slit loss by using the $r$-band DECaLS imaging to compute the ratio of the total galaxy flux to the fraction of flux captured by the slit. No dust corrections were applied to the SFRs. The SFRs are listed in Table~\ref{tab:gal_params}.

\begin{table*}
    \centering
    \begin{tabular}{lccccccccccc}
        \hline 
        Name & RA & DEC & $z_{\rm gal}$ &
        $D$ & {\Rv} &$D/R_{\rm vir}$ & $M_*$ & $M_{\rm H}$ & SFR & $i$ & $\Phi$  \\ 
        & (J2000) & (J2000)  &  & (kpc) & (kpc) & & ($M_\odot$) & ($M_\odot$) & ($M_\odot$~yr$^{-1}$) & (degree) & (degree)\\
        (1) & (2) & (3) & (4) & (5) & (6) & (7) & (8) & (9) & (10) & (11) & (12)\\
        \hline
            G1 & 00:33:40 & $-$00:55:22	 & 0.2124	 & 21.9	 & 110&0.20	 & 9.8	 &11.3  & 0.25$^{\ddagger}$& 81 & 10 \\
            G2 & 00:34:13 & $-$01:00:20	 & 0.2565	 & 30.4	 & 142&0.21	 & 10.3	 &11.6  & 0.10$^{\dagger}$& 85 & 29\\
            G3 & 01:01:56 & $-$08:44:09	 & 0.1583	 & 28.4	 & 91 &0.31	 & 9.4	 &11.1  & 0.31$^{\dagger}$& 67  & 35\\
            G4 & 12:13:09 & $+$14:08:37	 & 0.2903	 & 31.0	 & 152&0.20	 & 10.4	 &11.7  & 0.56$^{\ddagger}$& 73 & 9\\
            G5 & 08:23:41 & $+$07:47:51	 & 0.1866	 & 37.5	 & 171&0.22	 & 10.6	 &11.9  & 0.39$^{\ddagger}$& 60 & 25\\
            G6 & 13:27:57 & $+$10:11:36	 & 0.2554	 & 25.3	 & 106&0.24	 & 9.7	 &11.3  & 1.17$^{\ddagger}$& 63 & 1\\
            G7 & 12:16:41 & $+$07:12:24	 & 0.2366	 & 13.0	 & 106&0.12	 & 9.7	 &11.3  & 0.67$^{\ddagger}$& 61 & 19\\
            G8 & 21:29:39 & $-$06:37:59	 & 0.2780	 & 27.7	 & 117&0.24	 & 9.9	 &11.4  & 0.72$^{\ddagger}$& 66 & 9\\
            G9 & 23:49:49 & $+$00:35:42	 & 0.2782	 & 31.6	 & 163&0.19	 & 10.5	 &11.8  & 0.59$^{\ddagger}$& 78 & 6\\
        \hline 
    \end{tabular}
    \caption[]{
    Columns are: (1) Galaxy name.  
    (2) Galaxy right ascension.  
    (3) Galaxy declination.
    (4) Galaxy emission-line redshift.
    (5) Impact parameter adopted from \citet{huang21}.
    (6) Halo virial radius calculated following the formalism of \citet{bryan98}. 
    (7) Ratio of the impact parameter and virial radius.
    (8) Logarithm of the stellar mass adopted from \citet{huang21}.
    (9) Logarithm of the halo mass computed using the stellar-to-halo mass relation from  \citet{girelli20}.
    (10) Star formation rate from {\OII} ($\dagger$) or {\Ha} ($\ddagger$) emission.  
    (11) Galaxy inclination angle.
    (12) Galaxy azimuthal angle where zero degrees is along the galaxy major axis.
    }
\label{tab:gal_params}
\end{table*}

\subsection{Galaxy Morphologies}
The galaxy properties were measured using GALFIT \citep{peng10} on DECaLS \citep{dey19} $r-$band images. Point spread functions were modelled using unsaturated, nearby stars in the field. Galaxies were modelled in the $r-$band image. The galaxy inclination was computed from the GALFIT $b/a\equiv q$ values using $\cos^2(i) = (q^2 - q_0^2)/(1 - q_0^2)$, where $q_0\sim0.1$ for late type spirals \citep[e.g,][]{guthrie92,favaro25}. The galaxies are modestly inclined ($60\leq i \leq 85$  degrees, with a mean of $i=70\pm9$ degrees) and have their major axes aligned with the quasar sightlines (1 $\leq \Phi \leq $ 35 degrees, with a mean of $\Phi=15\pm12$ degrees). The galaxy properties are shown in Table~\ref{tab:gal_params}. Given the assumption of $q_0$, which can range from $q_0=0.1-0.2$ \citep[see summary from][]{favaro25} and the low spatial resolution images, the errors in the inclination values are of the order of ten degrees, while the azimuthal angle errors are of the order of five degrees. The galaxy orientation is not directly used in this study, but is used to emphasise that our galaxies are near edge-on with the quasar near their major axis. We are in the process of obtaining higher spatial resolution images to further address in the future how the CGM behaves relative to the disks of galaxies. 

\section{Results}
\label{sec:results}

In the next subsections, we present the properties of the sample, the kinematic comparison between the galaxy ISM and multi-phase CGM detected in absorption, and general statistical kinematic trends. 

\subsection{Sample Properties}

Figure~\ref{fig:EWD} shows the DECaLS \citep{dey19} $grz$ colour images of the nine galaxies. The quasars are the bright blue point sources and the targeted galaxies are the largest galaxy next to them. The unoriginal nicknames for each of these galaxies are labelled in the upper-corner of the figure and are used throughout the manuscript. All of these galaxies are found to be isolated to at least with 130~kpc, based on the spectroscopic survey of \citet{chen10a} and \citet{huang21}. G1 has a bright nearby galaxy, but it is at an unrelated redshift.

The galaxies have {\Ha} or {\OII} derived star-formation rates ranging between $0.1\leq$SFR$\leq 1.1$~M$_{\odot}$~yr$^{-1}$.
The galaxies are edge-on, with a mean inclination angle of $70\pm9$ degrees and a range of $60\leq i\leq 85$ degrees. The quasar sightlines are located at azimuthal angles ranging between 1 $\leq \Phi \leq $ 35 degrees, with a mean of $\Phi=15\pm12$ degrees.

Figure~\ref{fig:EWD} further shows the {\MgII} absorption strength for the 9 galaxies as blue squares ($\bluesquare$). For reference we also show the {\magiicat} \citep{magiicat2,magiicat1} catalogue along with their fit to the data. Our sample shows the similar trend of decreased {\MgII} rest-frame equivalent width with increasing projected galaxy-quasar separation ($D$). The sample is consistent within the scatter of the {\magiicat} data. Some of the larger impact parameter systems seem to reside below the {\magiicat} data, but this may likely be attributed our sample being major axis systems, where weaker absorption tends to reside \citep{bordoloi11,lan18}. We will present cloud properties of all the multi-phase absorption in future works, which will focus on the detailed analysis of the absorption properties.

\subsection{Galaxy Gas--CGM Kinematic Comparison}
\begin{figure*}
	\includegraphics[width=\textwidth]{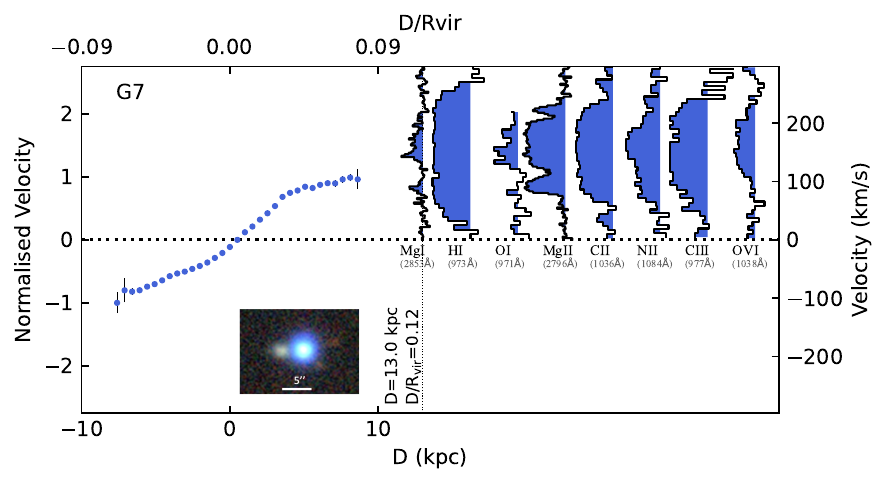}
    \caption{Kinematic comparison between the galaxy ISM rotation curve for G7 and the multi-phase CGM absorption, with the quasar field image for reference. Blue circles represent the galaxy's rotation curve, with rotation toward the quasar in the upper right quadrant. The y-axes show line of sight velocity relative to the galaxy systemic (right) and normalized by peak galaxy rotation velocity (left). The top and bottom axes indicate the impact parameter ($D$) and $D/\Rv$, respectively, with the quasar distance labelled in the figure. Absorption profiles at the quasar sightline for ions of increasing ionisation potential are shown, offset for clarity. This figure allows a direct comparison of galaxy and CGM kinematics, demonstrating that the multi-phase CGM aligns with the galaxy's rotation direction.}
    \label{fig:G7kine}
\end{figure*}

\begin{figure*}[!h]
	\includegraphics[width=\textwidth]{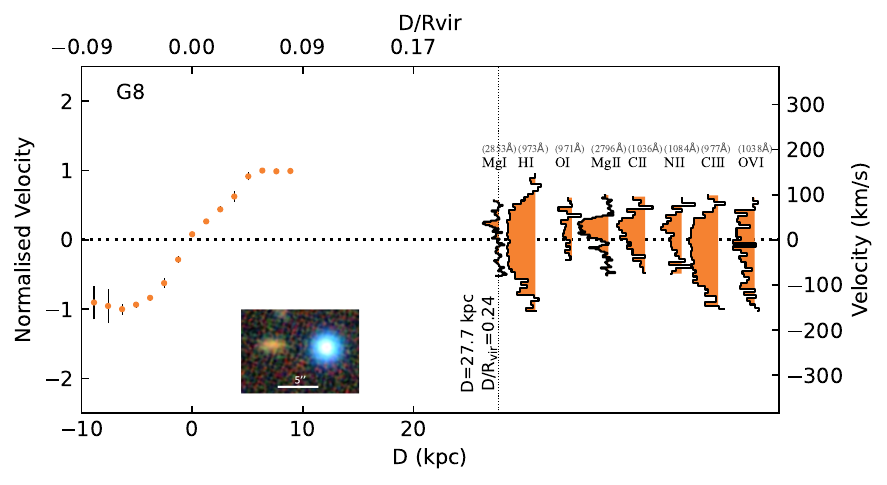}
    \caption{Same as Figure~\ref{fig:G7kine} but for G8, having a $D/\Rv = 0.24$, with the rotation toward the quasar in the upper right quadrant. While the strongest absorption component from {\MgI} and {\MgII} are consistent with the galaxy’s rotation direction, but slower than the maximum rotation speed, absorption in the opposite direction is also observed, particularly for {\CIII} and {\OVI}.}
    \label{fig:G8kine}
\end{figure*}

Here we explore how the galaxy's ISM rotation direction and speed corresponds to the kinematics of the CGM found along the quasar sightlines. Figure~\ref{fig:G7kine} shows an example of this kinematic comparison for galaxy G7. The $grz$ colour image of G7 and the quasar is shown in the lower left of the figure. The blue circle data points show the rotation curve of G7 with the y-axes showing both the line-of-sight velocity in {\kms} (right-side) as well as the peak rotation-normalised velocity (left side). The x-axes show the projected distance from the galaxy towards the quasar in kpc ($D$; lower axis) and in virial radius-normalised distance ($D/R_{\rm vir}$; top axis). The halo virial radius are calculated following the formalism of \citet{bryan98}. The direction towards the quasar is in the positive x-axis direction and the direction of galaxy rotation towards the quasar is in the positive y-axis direction above the horizontal dashed line; e.g., the top half of the panel. This relative galaxy-quasar orientation applies to all figures for all the other galaxies  (Figures~\ref{fig:G8kine},~\ref{fig:G1kine},~\ref{fig:G2kine},~\ref{fig:G3kine},~\ref{fig:G4kine}~\ref{fig:G5kine},~\ref{fig:G6kine},~\ref{fig:G9kine}).  The vertical dotted line shows the projected distance of the quasar, which is also labelled. At the projected distance of the quasar, the spectra of individual CGM absorption lines are shown vertically in order of increasing ionisation potential (the ionisation energy required to remove an electron from the neutral atom to create the ion) and are offset for clarity. Not all of the absorption lines detected in the quasar spectra are shown, but we have chosen to display a subset that are common for each galaxy ({\MgI}, {\HI}, {\OI}, {\MgII}, {\CII}, {\NII}, {\CIII} and {\OVI}). {\MgI} and {\MgII} are from UVES data and have higher spectral resolution ($\sim6$~\kms) compared to the other ions obtained with COS ($\sim20$~\kms). Example galaxies G7 and G8 are shown in Figures~\ref{fig:G7kine}$-$\ref{fig:G8kine}, with the remaining galaxies shown in Figures~\ref{fig:G1kine}$-$\ref{fig:G9kine}.

Directly comparing the ISM to the CGM, Figure~\ref{fig:G7kine} shows galaxy G7 rotating at $100$~{\kms} in the direction of the quasar. G7 is located at $D=13$~kpc, or $D/R_{\rm vir}=0.12$, from the quasar sightline. The CGM is detected in all ions displayed. Given the full Lyman series is covered in the COS spectra, from {\Lyb} blueward, we select a higher-order Lyman transition to display (Ly$\gamma$) to further show velocities that the larger optical depths reside at and the kinematic spread of the {\HI}. The CGM is found to be rotating in the same direction as the galaxy, regardless of the ion selected, and all of the CGM is found to one side of the galaxy systemic velocity. The {\MgI} absorption exhibits two main velocity components; one slightly lower than the rotation speed of the galaxy and another component at velocities higher than the rotation speed. Future ionisation modelling work will aim to determine whether the physical properties of those different components differ or not. The {\MgII} absorption is broader with additional velocity components and has a velocity range of $50-250$~{\kms}. This is similar to what is found for other ions, with {\CII} and {\CIII} having larger velocity spreads and are all in the same direction of rotation of the galaxy.

Figure~\ref{fig:G8kine} shows the higher impact parameter galaxy G8, located at $D=28$~kpc ($D/R_{\rm vir}=0.24$) from the quasar sightline, rotating at $150${\kms} in the direction of the quasar. Again, the direction of rotation towards the quasar is in the upper half of the figure. 

Although the quasar sightline is at a larger impact parameter compared to G7, all the ions are detected. However, the CGM absorption associated with G8 is weaker than G7, which is consistent with the general anti-correlation between equivalent width and impact parameter. 

For G8, we find that velocities of {\MgI} and {\OI} are in the same direction of the galaxy rotation with velocities lower than the galaxy rotation speed. The {\MgII}, {\CII}, and {\NII} have at least two components; the largest optical depth component is consistent with the rotation direction of the galaxy towards the quasar sightline, while the lower column density component is inconsistent with rotation. The {\HI}, {\CIII} and {\OVI} span the systemic velocity on both sides, with a preference to being opposite to the direction of rotation. Thus, in this example, while the highest optical depth gas is consistent with the rotation direction of the galaxy, there is some absorption that is disconnected with the rotation of the galaxy and this is most apparent in the high ions. 

Similar trends are seen for the remaining galaxies shown in Figures~\ref{fig:G1kine}$-$\ref{fig:G9kine} in the Appendix with a discussion of each individual galaxy and CGM. In the remainder of the paper, global kinematic trends will be discussed.

\begin{figure*}
\includegraphics[width=\textwidth]{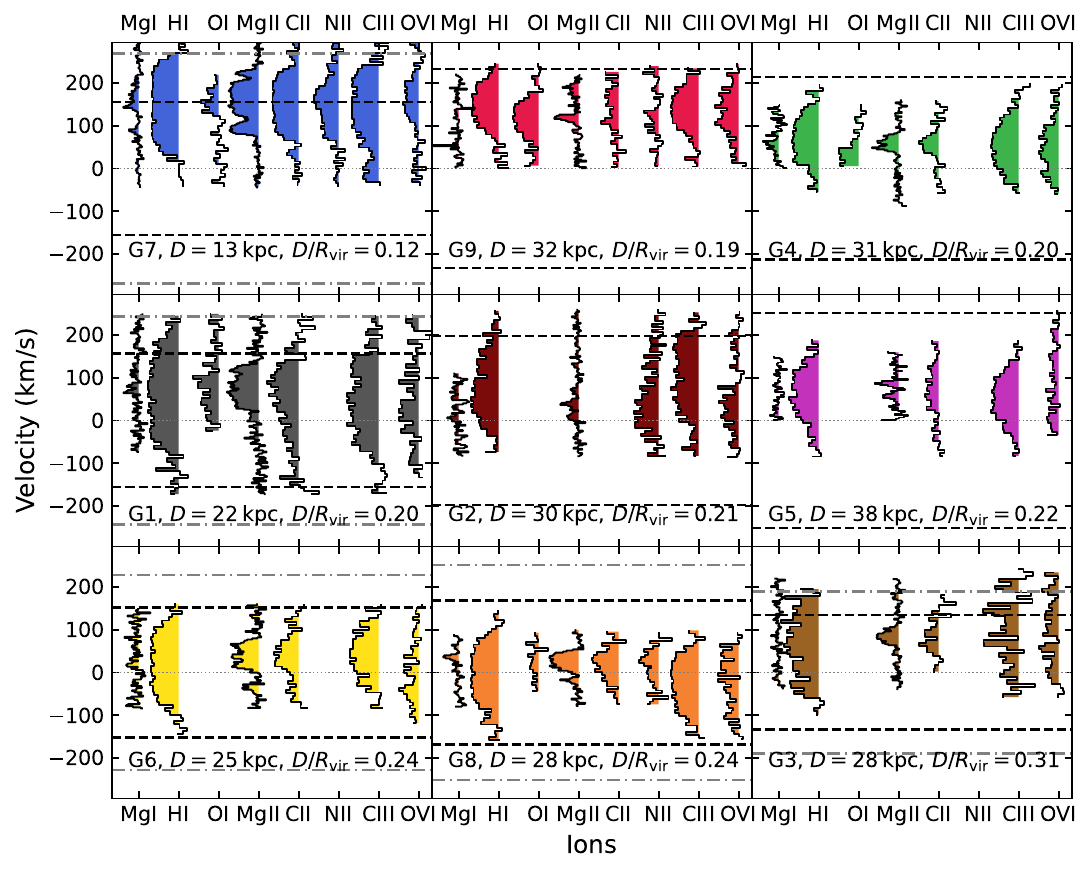}
    \caption{Kinematic comparison of the multi-phase CGM absorption relative to the escape velocity for galaxies G1 to G9. Galaxies are ordered in increasing $D/${\Rv} from left to right, top to bottom. Absorption features for each galaxy are shown as histograms, with the line-of-sight velocity on the y-axis and the ion name along the x-axis indicates the different ions. The black dashed lines represent the escape velocity at {\Rv} and the grey dashed lines represent the escape velocity at projected distance $D$, for each galaxy. Galaxies in the figure without grey dashed lines have escape velocities at $D$ exceeding the plotted range of $\pm320$~{\kms} (G2, G4, G5, and G9). The escape velocity lines allow for a comparison of the multi-phase CGM kinematics with the galaxy's gravitational potential, where the gas may be gravitationally bound or exceed the escape velocity. For the escape velocity at {\Rv}, 8/9 systems show all of the low ionisation gas being bound to the halo while higher ionisation gas is bound to the halo for 6/9 galaxies. All of the CGM is bound at the projected distance. }
    \label{fig:vescape}
\end{figure*}

\subsection{Escaping or bound?}

While studies have shown that the majority of multi-phase CGM appears to be gravitationally bound to galaxies, there are a small fraction of absorption systems that do have optical depth-weighted velocity centroids, or higher velocity absorption wings, exceeding the escape velocity \citep[e.g.,][]{tumlinson11, bordoloi14,kacprzak19kine,huang21,dutta25}. Here we explore the kinematics of major axis absorption relative to the escape velocity.

Figure~\ref{fig:vescape} shows the escape velocities for each galaxy, calculated both at the virial radius (black dashed lines) and at the impact parameter distance (grey dashed-dotted lines). The nine panels show different galaxies, labelled G1 to G9, with galaxies ordered in increasing $D/${\Rv} from left to right, top to bottom. Galaxies in the figure without grey dashed lines have escape velocities at $D$ exceeding the plotted range of $\pm320$~{\kms} (G2, G4, G5, and G9). The vertical axis represents the line of sight velocity in {\kms}, and the x-axis corresponds to ions observed. Positive velocities are in the direction of galaxy rotation towards the quasar.  

First we examine the CGM relative to the escape velocity at the virial radius. Eight out of nine galaxies have all their low ionisation CGM ({\MgI} to {\CII}) below the escape velocity. The 
one exception is G7, which has a large fraction of the absorption above the escape velocity (this is also the lowest impact parameter system with $D=13$~kpc, $D/$\Rv$=0.12$). For the higher ionisation CGM ({\CIII} and {\OVI}) and {\HI}, three out of nine galaxies (G7, G1 and G3) have absorption velocities slightly above the escape velocity. Thus except for G7, 8/9  (89\%) all of the low ionisation gas is bound to the halo while higher ionisation gas is bound to the halo for 6/9 (67\%) galaxies.  

We further compare the CGM to the escape velocity at $D$ for each galaxy, which is indicated by the grey dashed line. If this gas is in an extended disk or accreting co-planar filament, then the assumption of the CGM being near or at $D$ is not unrealistic.  Assuming that the CGM is at or near $D$ within the halo, then Figure~\ref{fig:vescape} demonstrates that all of the gas is bound to the galaxy in all cases and for all gas phases. Thus, it seems likely that all of the detected CGM absorbers are bound to the galaxy in each case.

\begin{figure*}
\includegraphics[width=\textwidth]{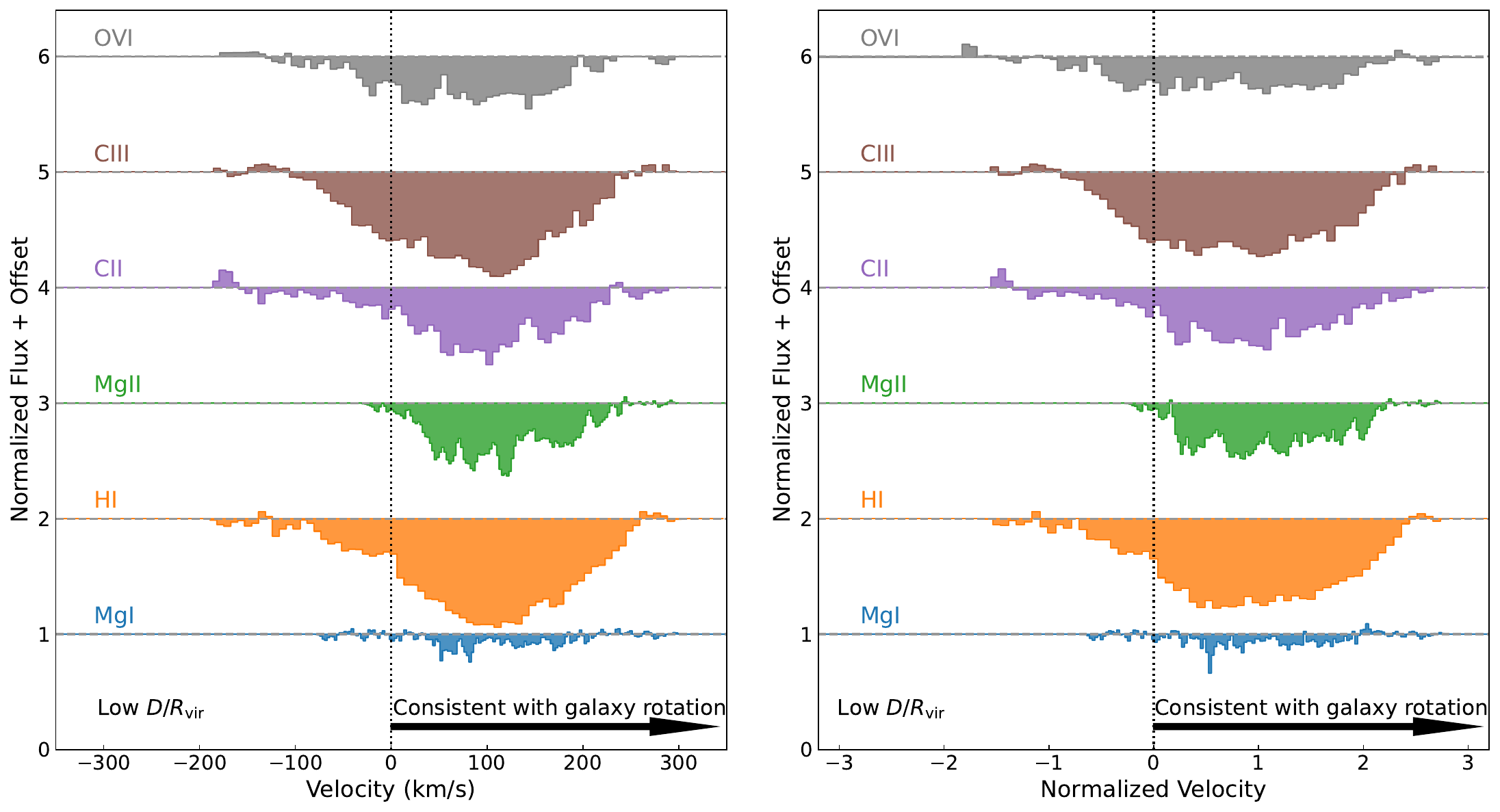}
    \caption{Stacked {\MgI}, {\HI}, {\MgII}, {\CII}, {\CIII} and {\OVI} absorption lines for the low $D/${\Rv} bin ($0.12\leq$$D/R_{\rm vir}$$\leq0.20$). The spectra are offset along the y-axis in order to compare their kinematic structure. The x-axis shows (left) line-of-sight velocity and (right) normalised velocities relative to the maximum galaxy rotation speed. Positive velocities align with the direction of galaxy rotation towards the quasar sightline.  At low $D/{\Rv}$, the {\MgI} and {\MgII} are fully aligned with the direction of rotation of the galaxy, while {\HI}, {\CII}, {\CIII} and {\OVI} have a small fraction absorption in the opposite direction of rotation. }
    \label{fig:stack_low}
\end{figure*}

\begin{figure*}
\includegraphics[width=\textwidth]{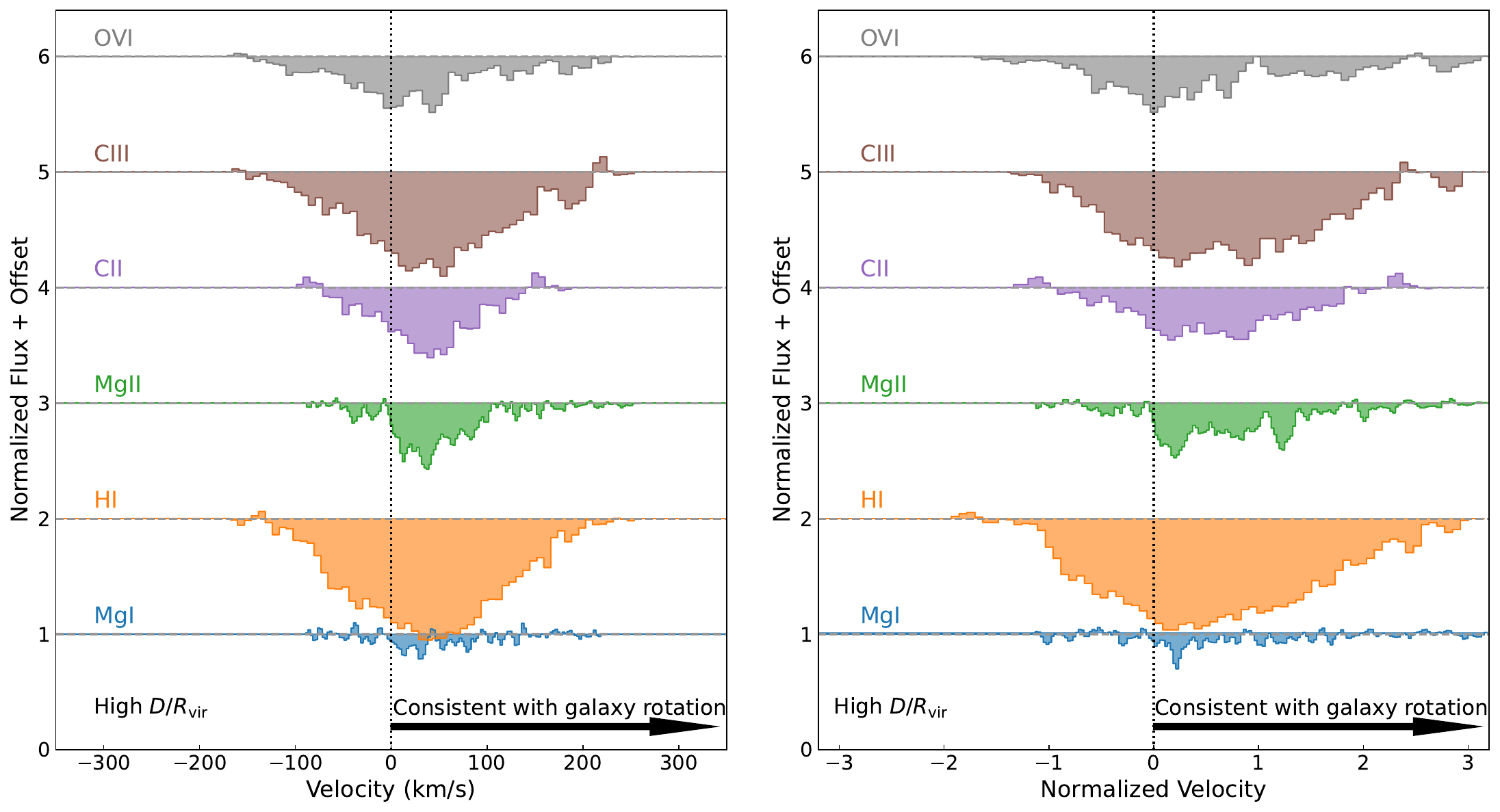}
    \caption{Same as Figure~\ref{fig:stack_low} except the stacked spectra are now plotted for the high $D/${\Rv} bin ($0.21\leq$$D/R_{\rm vir}$$\leq0.31$). In contrast to the low $D/${\Rv} subsample, {\MgII} and possibly {\MgI} have a small fraction of absorption in the opposite direction of rotation. Furthermore, most of their optical depth is located at lower velocities closer to systemic. {\HI}, {\CIII} and {\OVI} exhibit a broad single absorption profile and have more absorption on the opposite side of the galaxy rotation direction.}
    \label{fig:stack_high}
\end{figure*}

\begin{figure*}
\centering
\begin{subfigure}{0.49\linewidth}
    \centering
    \includegraphics[width=\linewidth]{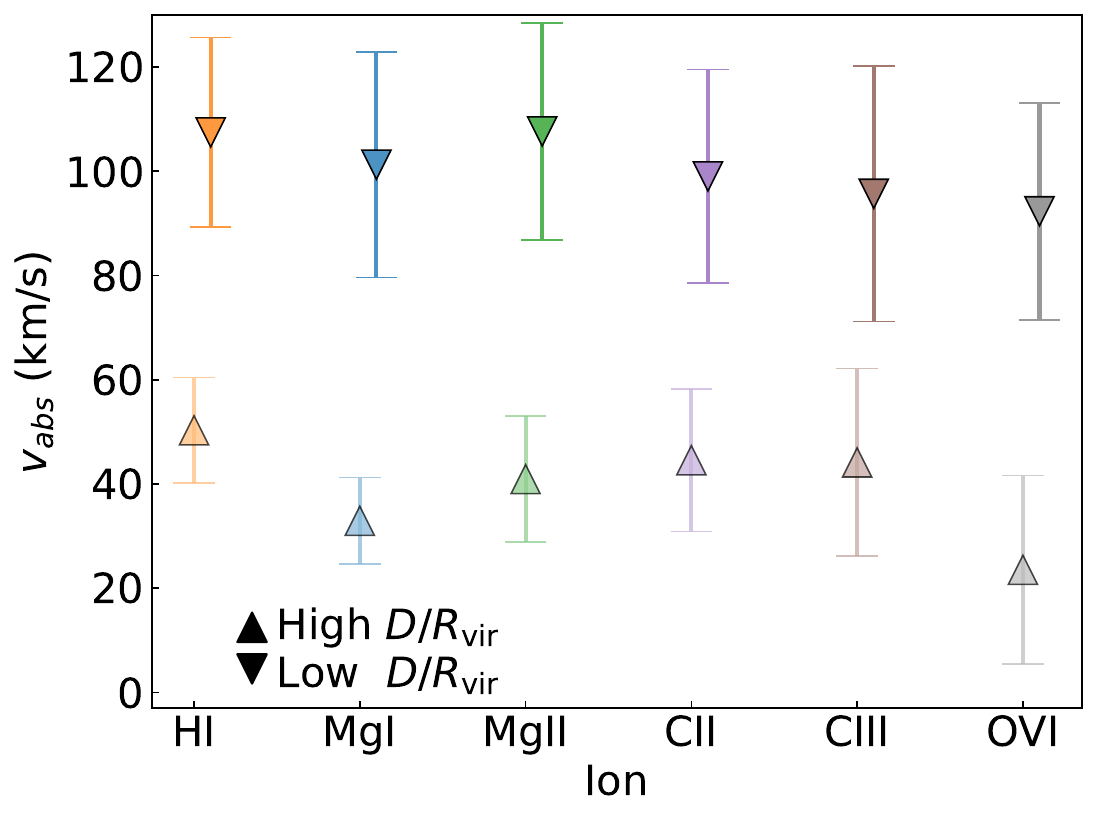}
\end{subfigure}
\hfill
\begin{subfigure}{0.49\linewidth}
    \includegraphics[width=\linewidth]{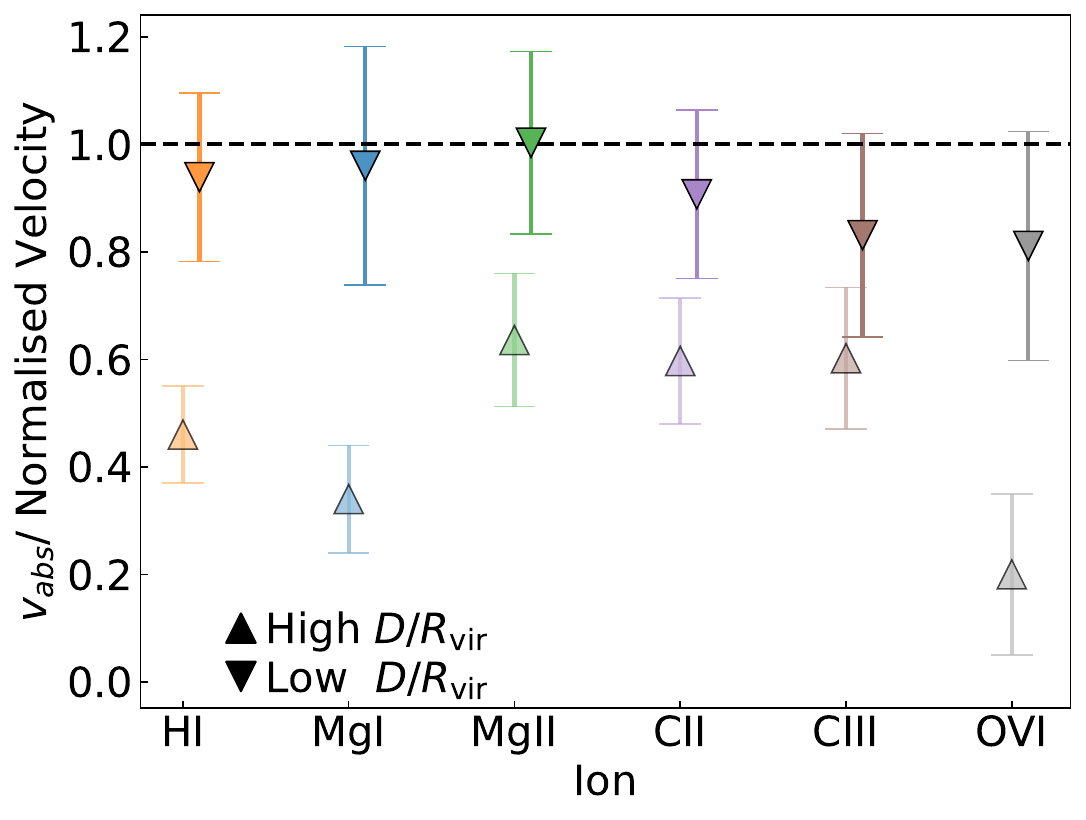}
 \end{subfigure}
\caption{(Left) Optical depth weighted median velocity, $v_{\rm abs}$, of the bootstrapped-stacked spectra for each ion relative to the galaxy systemic velocity ($v=0$~{\kms}), where the positive values indicate gas moving in the direction of the galaxy rotation, for low and high $D/R_{\rm vir}$ bins. At low $D/{\Rv}$, the optical depth weighted velocity for all ions is consistent with 110~{\kms} with no apparent shift between ions. For high $D/{\Rv}$, the  optical depth weighted velocity decreases by a factor of up to 2.75 and {\OVI} exhibits the largest shift. (Right) Optical depth weighted median velocity of each ion relative to the galaxy maximum normalised velocity, where the dashed line indicates the maximum rotation speed. The optical depth weighted median velocity at lower $D/R_{\rm vir}$ is consistent with the maximum rotation speed of the galaxy. At higher $D/R_{\rm vir}$, the optical depth weighted median velocity is lower than the rotation speed of the galaxy, with the largest difference found for {\OVI}.}
\label{fig:dv}
\end{figure*}
 
\begin{figure*}
\centering
\begin{subfigure}{0.49\linewidth}
    \centering
    \includegraphics[width=\linewidth]{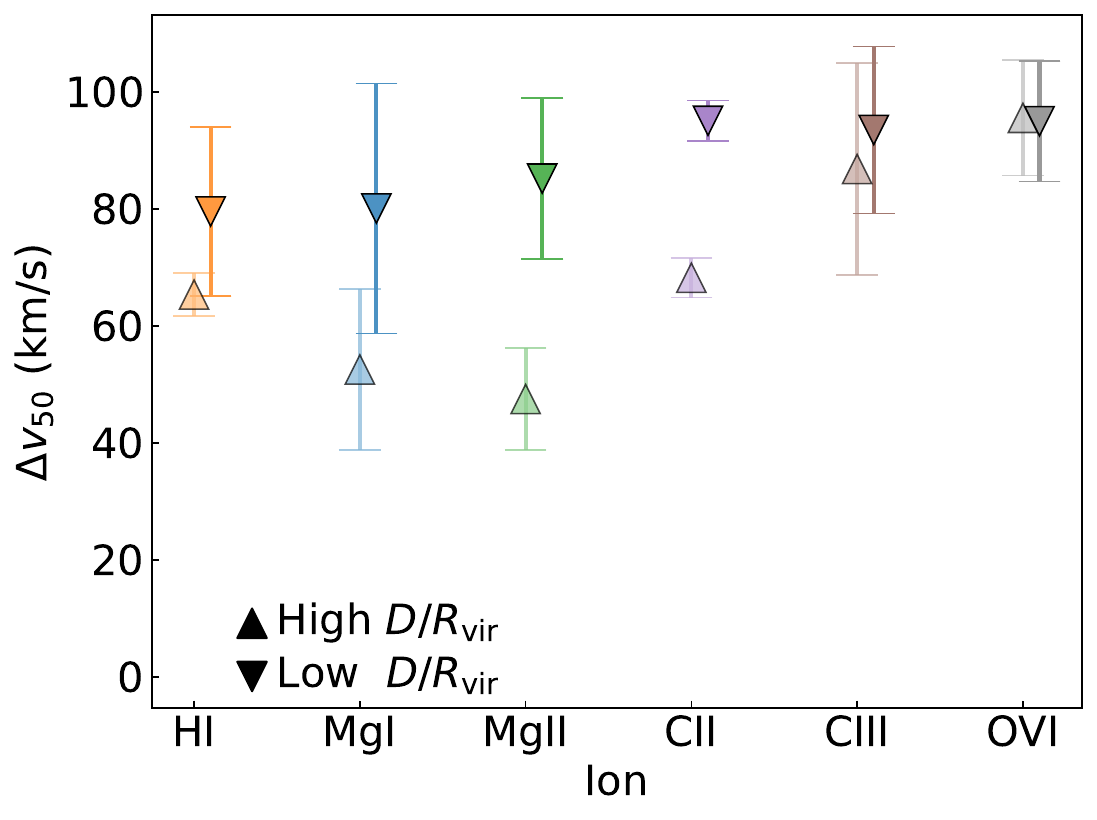}
\end{subfigure}
\hfill
\begin{subfigure}{0.49\linewidth}
    \includegraphics[width=\linewidth]{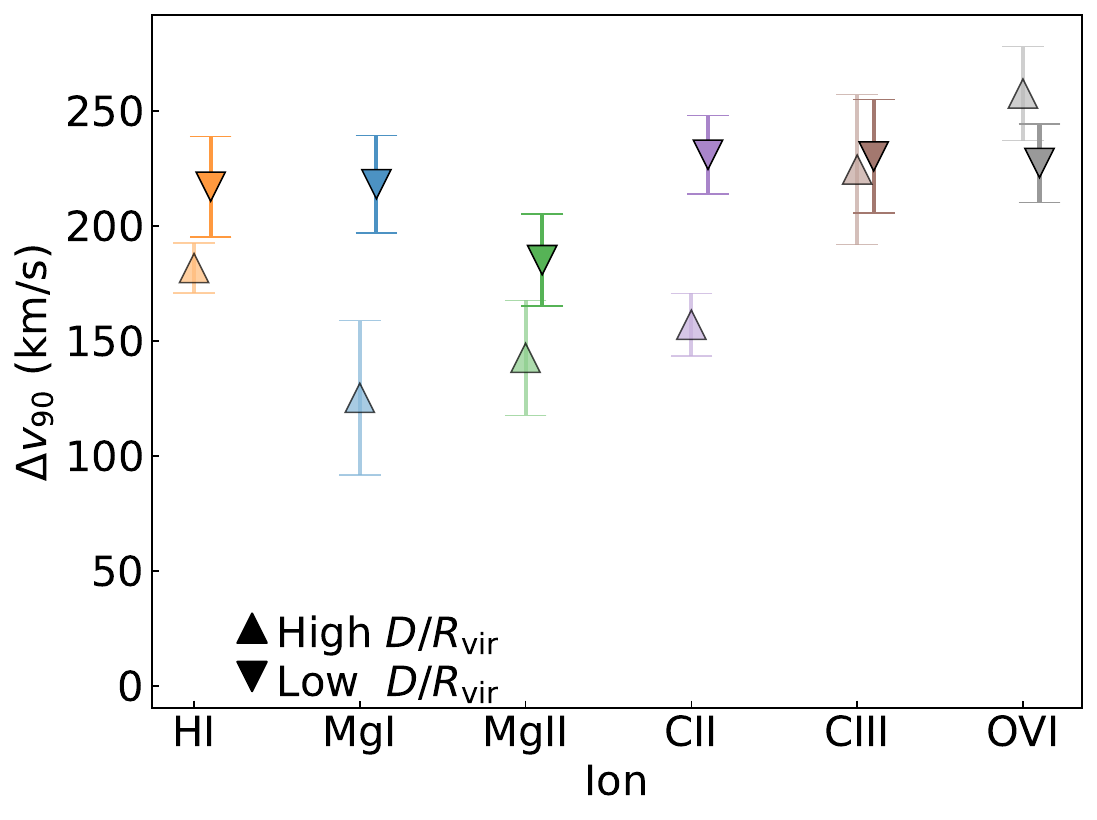}
 \end{subfigure}
\caption{Velocity widths corresponding to 50\%  ($\Delta v_{50}$) (left) and 90\% ($\Delta v_{90}$) (right) of the total optical depth for the bootstrapped-stacked spectra for each ion in the low $D/{\Rv}$ and high $D/{\Rv}$ bins. Larger $\Delta v_{50}$ values are found closer to the galaxies and for more highly ionised gas, with increasing velocity width for the higher ionisation ions. The increase $\Delta v_{50}$ is shallow for gas close to galaxies (low $D/{\Rv}$) and and more steep for gas further from galaxies (high $D/{\Rv}$). For the most highly ionised gas  traced by  {\CIII} and {\OVI}, there is no difference in $\Delta v_{50}$ with $D/{\Rv}$. For $\Delta v_{90}$, there is a possible shallow increase in velocity from low to high ionisation gas for low $D/{\Rv}$, whereas there is a steeper increase for high $D/{\Rv}$. Again, there is no change in $\Delta v_{90}$ for {\CIII} and {\OVI} with $D/{\Rv}$.}
\label{fig:dv5090}
\end{figure*}


\subsection{Co-rotation Statistics as a Function of $D/R_{\rm vir}$}

\subsubsection{Stacked spectra}

We investigate the general kinematic trends as a function of $D/${\Rv}. We split the sample into lower $D/${\Rv} (G1, G4, G7, G9) and higher $D/${\Rv} (G2, G3, G5, G6, G8) subsamples, having mean and standard error in the mean values of $D/${\Rv}$=0.18\pm0.02$ ($0.12\leq$$D/R_{\rm vir}$$\leq0.20$) and $D/${\Rv}$=0.24\pm0.02$ ($0.21\leq$$D/R_{\rm vir}$$\leq0.31$), respectively. Throughout the paper, the low-$D/R_{\rm vir}$ results are shown at full colour opacity, whereas the high-$D/R_{\rm vir}$ results are shown in the same colours but with reduced opacity (light shading) to aid comparison.

Figure~\ref{fig:stack_low} shows the stacked spectra of the {\MgI}, {\HI}, {\MgII}, {\CII}, {\CIII} and {\OVI} absorption lines for the low $D/${\Rv} sample. The spectra are arbitrarily offset from each other along the y-axis in order to compare their kinematic structure. Positive velocities align with the galaxy's rotation direction, where the left panel shows the line-of-sight velocity and the right panel shows velocities relative to the maximum galaxy rotation speed. In this latter case, each spectrum was normalized to its own galaxy rotation velocity and then combined afterwards, making for a fairer comparison between objects. At low $D/{\Rv}$, the {\MgI} and {\MgII} are fully aligned with the direction of rotation of the galaxy, while spanning up to two times the maximum rotation speed of the galaxy. 

It appears as if HI comprises multiple components based on the low ionisation absorption structure, with the strongest component centred at $\sim100$~{\kms} (or a normalized velocity of unity), while the weaker one is in the opposite direction, counter-rotating at $-50$~{\kms} (or a normalized velocity of 0.25). However, another possible scenario is that there is a broader, weaker component at the systemic velocity ($\sim 0$~{\kms}) that is overlapped by the stronger co-rotating components. This similar spectral shape is also seen for {\CII}, {\CIII} and {\OVI}, as these ions have a minority of their absorption in the opposite direction of rotation. 

Figure~\ref{fig:stack_high} is the same as Figure~\ref{fig:stack_low}, except the data are for the galaxies at larger $D/{\Rv}$. In contrast to the low $D/{\Rv}$ sample, {\MgII} and possibly {\MgI} have a small fraction of absorption in the opposite direction of rotation. Furthermore, most of their optical depth is located at lower velocities closer to systemic. {\HI} exhibits a broad single absorption profile centred around 50~{\kms} and has more absorption on the opposite side of the galaxy rotation direction compared to the low $D/\Rv$ sample.  {\CIII} and {\OVI} follow the same kinematic spread and centroid trends as the {\HI}. 

Overall, we find that most absorption is consistent with the direction of rotation of the galaxy along the quasar sightline. However, sightlines at larger impact parameters tend to be less aligned with the direction of rotation, and this is most apparent for {\HI}, {\CIII} and {\OVI}. We next quantify the absorption profile centroids and widths in Section~\ref{sec:dv} and the absorption co-rotation fraction in Section~\ref{sec:co-rot}.

\subsubsection{Velocity centroids and widths}
\label{sec:dv}
We compute the optical depth weighted velocity centroids and the velocity widths corresponding to 50\% ($\Delta v_{50}$) and 90\% ($\Delta v_{90}$) of the total optical depth for each ion in the stacked low and high $D/R_{\rm vir}$ bins. To compute the uncertainties in our measurements and to account for sample variations, we employed a bootstrap resampling approach. We generated 1000 resampled stacked spectra for each of the low and high $D/R_{\rm vir}$ bins, then measured the velocity centroid and velocity width of each stacked spectrum for each ion. From the resulting bootstrap distributions, we computed the median and adopted the 16th and 84th percentiles of the bootstrap distributions as the $1\sigma$ uncertainties.

hows the optical depth weighted velocity (left) and the optical depth weighted velocity normalized by each galaxy’s maximum rotation speed (right) for each ion. Low $D/R_{\rm vir}$ bins are shown as solid colour downward pointing triangles ($\shadeddowntriangle$), and high $D/R_{\rm vir}$ bins as lightly shaded upward pointing triangles ($\shadeduptriangle$). All values and uncertainties are derived from our bootstrap analysis.

Figure~\ref{fig:dv} shows the optical depth weighted velocity (left) and the optical depth weighted normalized velocity normalized by each galaxy’s maximum rotation speed (right) for each ion. Low $D/R_{\rm vir}$ bins are shown as solid colour downward pointing triangles ($\shadeddowntriangle$), and high $D/R_{\rm vir}$ bins as lightly shaded upward pointing triangles ($\shadeduptriangle$). All values and uncertainties are derived from our bootstrap analysis. At low $D/{\Rv}$, the optical depth weighted velocity for all ions is consistent with 110~{\kms} with no apparent shift between ions. For high $D/${\Rv}, the  optical depth weighted velocity decreases by a factor of 2.75 to 40~{\kms}. {\OVI} is found at a lower velocity of $23\pm18$~{\kms}, which is a decrease of a factor of 4.8 compared to {\OVI} at low $D/{\Rv}$. 

Figure~\ref{fig:dv} (right) shows the the optical depth weighted normalized velocity with the dashed line indicating the maximum normalized rotation velocity of the galaxies. At low $D/{\Rv}$, the bootstrapped optical depth weighted velocity for all ions is consistent with the rotation speed of the galaxy, exhibiting the strongest connection between the galaxy rotation and the CGM absorption, although {\CIII} and {\OVI} could potentially be at lower velocities. For high $D/{\Rv}$, the normalised velocity decreases to $\sim0.6$, within {\OVI} having the lowest value at $0.20\pm0.15$. Thus, overall, we find that the kinematic connection between the CGM and galaxy can drop as much as a factor 2.75 in velocity, or by a factor of 1.7 when considering the normalised velocity between low and high $D/{\Rv}$.  

While the optical depth weighted velocity is informative on the alignment of the bulk of the CGM with the galaxy, the $\Delta v_{50}$ and $\Delta v_{90}$ better quantify the kinematic spread of the absorption. Figure~\ref{fig:dv5090} shows the $\Delta v_{50}$ (left) and $\Delta v_{90}$ (right) for the bootstrapped stacked spectra as a function of $D/{\Rv}$ and ionisation. Both {\MgII} and {\CII} point to the larger $\Delta v_{50}$ closer to the galaxies. At low $D/{\Rv}$, the {\MgII} $\Delta v_{50}=85$~{\kms} while at high $D/{\Rv}$ its $\Delta v_{50}=50$~{\kms}, which is a factor of 1.8 decrease from low to high $D/{\Rv}$. It also appears that the $\Delta v_{50}$ increases with increasing ionisation {\MgI} to {\OVI}, indicating the higher ionisation gas has larger velocity dispersions. This increase is strongest in the high $D/{\Rv}$ bin. We find no change for both {\CIII} and {\OVI} with distance, where $\Delta v_{50}$ remains large for these highest ionisation gas tracers.

Figure~\ref{fig:dv5090} shows the $\Delta v_{90}$ (right) for the bootstrapped stacked spectra as a function of $D/{\Rv}$ and ionisation.  $\Delta v_{90}$ exhibits a similar trend to that seen for $\Delta v_{50}$. $\Delta v_{90}$ appears to be broader at low $D/\Rv$ with a flat or shallow slope with increasing ionisation.  This slope becomes steeper at high $D/\Rv$, with no change for both {\CIII} and {\OVI} with distance.

Overall, we find that for all ions, the inner CGM more closely traces the galaxy’s rotation, yet exhibits larger
$\Delta v_{50}$ and $\Delta v_{90}$ values than gas located at greater distances. {\OVI} appears to show the most disconnect from the galaxy kinematics.

\begin{figure*}[hbt!]
\centering
\begin{subfigure}{0.49\linewidth}
    \centering
    \includegraphics[width=\linewidth]{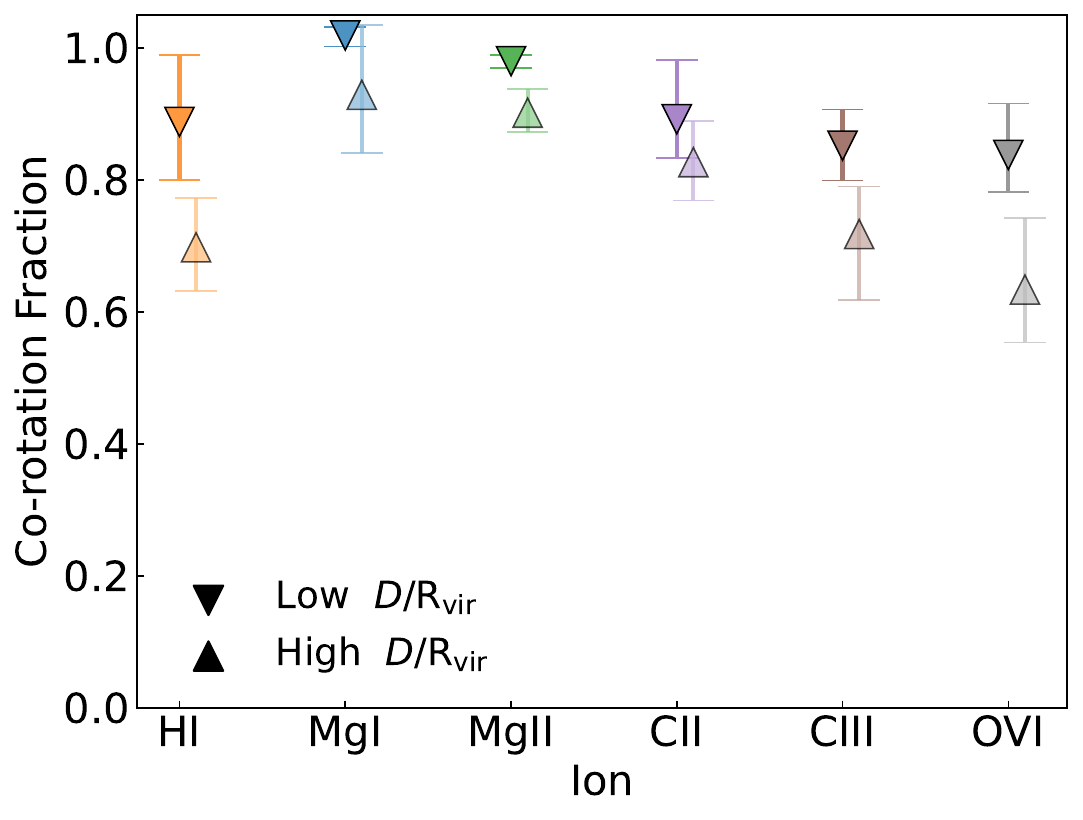}
\end{subfigure}
\hfill
\begin{subfigure}{0.49\linewidth}
    \includegraphics[width=\linewidth]{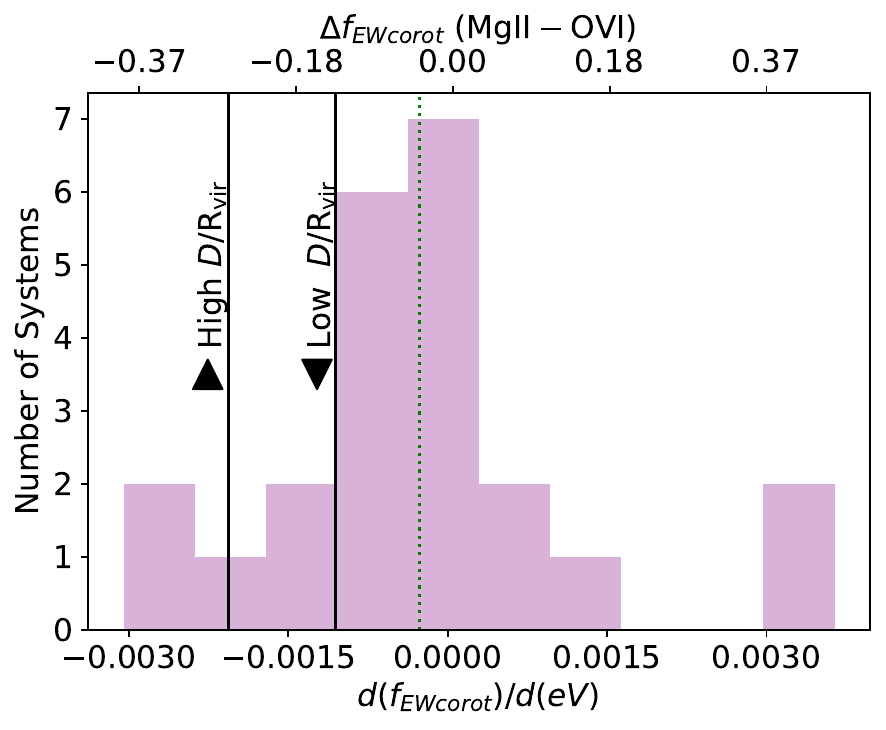}
 \end{subfigure}
\caption{(Left) Rest-frame equivalent width co-rotation fraction \citep[measurement adopted from][]{nateghi24GFI,nateghi24GFII} for both low and high $D/{\Rv}$ bins for {\HI} and the metal lines. A value of unity implies all of the absorption is consistent with the direction of rotation of the galaxy. The CGM becomes more disconnected from galaxy rotation for higher ionisation gas and with increasing $D/{\Rv}$; the largest change occurs for {\CIII}, {\OVI} and {\HI}. (Right) Adapted from \citet{nateghi24GFII}, showing their distribution of the change in the co-rotation fraction as a function of ionisation potential ({\slope}), with the top x-axis showing the difference between co-rotation fractions of {\MgII} and {\OVI}: \dfcorot. The green line represents the average of their sample, where the distribution shows that the majority of the systems that exhibit low-ionisation gas has a higher co-rotation fraction compared to the higher-ionisation phase. The black lines indicate the slope for the low and high $D/{\Rv}$ bins for the galaxies presented here.   }
\label{fig:corotions}
\end{figure*}

\subsubsection{Co-rotation fraction}
\label{sec:co-rot}
To further quantify how much absorption is consistent with the rotation of the galaxy, we employ the method developed by \citet{nateghi24GFI} and \citet{nateghi24GFII}. The equivalent width co-rotation fraction is defined as the fraction of the total equivalent width in the direction of the galaxy rotation such that:
\[
  f_{\rm EWcorot}
  \;=\;
  \frac{\displaystyle
    \int_{v_{\rm sys}}^{v_{\rm max}}
      \bigl[1 - F(v)\bigr]\,\mathrm{d}v}
  {\displaystyle
    \int_{v_{\rm min}}^{v_{\rm max}}
      \bigl[1 - F(v)\bigr]\,\mathrm{d}v}\;,
\]
where \(F(v)\) is the continuum‐normalized flux at velocity \(v\) relative to the systemic velocity \(v_{\rm sys}\), and \(v_{\rm min}\) and \(v_{\rm max}\) are the minimum and maximum velocities over which significant absorption is detected.  All absorption with \(v>v_{\rm sys}\) (i.e.\ sharing the sign of the galaxy’s rotation) is counted in the numerator, while the denominator includes the full velocity range of the line profile.  By construction, $0\le f_{\rm EWcorot} \le 1$, since any counter‐rotating absorption (\(v<v_{\rm sys}\)) contributes only to the denominator.  Thus \(f_{\rm EWcorot}=1\) indicates that 100 percent of the absorption lies on the co‐rotating side, and \(f_{\rm EWcorot}=0\) indicates none does.

Figure~\ref{fig:corotions} (left) shows the bootstrapped co-rotation fractions on the stacked absorption for the low and high $D/R_{\rm vir}$ bins. Again, to compute the 1~$\sigma$ uncertainties in our measurements and to account for sample variations, we generated 1000 resampled stacked spectra for each of the low and high $D/R_{\rm vir}$ bins. 

For low $D/R_{\rm vir}$, the co-rotation is above 0.8 for all ions. There is an anti-correlation between the co-rotation fraction and ionisation energy, where {\MgI} and {\MgII} are the highest and consistent with a unity co-rotation fraction, while {\OVI} has a co-rotation fraction of 0.84. This results in a $\Delta f_{\rm EWcorot}$(\MgII--\OVI) change of nearly 16\% from the cool to the warmer phases. 
For the high $D/R_{\rm vir}$ bin, the co-rotation fraction for each ion drops relative to the low $D/R_{\rm vir}$ bin. {\MgII} decreases from unity to 0.90 when going to larger $D/R_{\rm vir}$. The other ions show similar decreasing trends, with {\OVI} decreasing to 0.63. This results in a $\Delta f_{\rm EWcorot}$(\MgII--\OVI) change of 26\% from the cool to the warmer phase. 

We compare our $\Delta f_{\rm EWcorot}$(\MgII--\OVI) to those reported by \citet{nateghi24GFII}. Figure~\ref{fig:corotions} (right) shows the distribution of $\Delta f_{\rm EWcorot}$(\MgII--\OVI) of all the galaxies from \citet{nateghi24GFII} as a pink histogram. Their galaxies span a range of azimuthal and inclination angles. They reported that on average, galaxies tend to have a decreasing correlation (negative slope) of the co-rotation fraction with increasing ionisation, i.e., {\MgII} is more coupled to the galaxy rotation compared to {\OVI}. Here we find the same trend for our stacked sample and further note that $\Delta f_{\rm EWcorot}$(\MgII--\OVI) decreases with distance from the galaxy. Also shown in Figure~\ref{fig:corotions} is the slope measured for the low and high $D/R_{\rm vir}$ bins indicated by the vertical lines and labels. While low $D/R_{\rm vir}$ is more consistent with the mean of their distribution (vertical green dotted line), the high $D/R_{\rm vir}$ sample is consistent with the more extreme systems in their sample. This is consistent with the previous subsection where {\OVI} becomes less kinematically coupled to the galaxy at larger distances. 



\section{Discussion}
\label{sec:discussion}
Our results support a kinematic connection between the multi-phase CGM gas and host-galaxy disk rotation, with that coupling decreasing for higher-ionisation gas and at larger distances. In the low $\leq0.2 R_{\rm vir}$ regime {\MgI}, {\MgII}, {\CII}, {\CIII} and {\OVI} absorption are predominantly consistent with a co-rotation model, matching galaxy rotation speeds, tracing an extended, rotating gaseous structure beyond the ISM. At larger $>0.2 R_{\rm vir}$ distances, this connection weakens for all ions with the largest change noted for {\OVI} absorption, having a stacked optical depth weighted median velocity 4.8 times slower than the rotation speed of the galaxies. On the other hand, {\MgII} absorption is only a factor of 2.75 slower than the rotation speed of the galaxies. This indicates that while the low-ionisation CGM remains more kinematically coupled to the galactic disk, with some decrease with distance, the high-ionisation phase decouples faster with $D/{\Rv}$. It must be noted that the optical depth weighted median absorption velocity for all ions is still consistent with the rotation direction of the galaxy, indicating that most of the multi-phase gas is consistent with a co-rotation/accretion model. However, for {\OVI} and {\HI} we find larger fractions of absorption at velocities in the opposite direction of rotation.

We observe a pronounced kinematic transition in the CGM over a narrow range of $D/\Rv$ from 0.1 to 0.3.  \citet{magiicat3} demonstrated that the mean {\MgII}~$\lambda2796$ equivalent width is essentially invariant with halo mass for ${D/R_{\rm vir}\lesssim0.3}$.
In their {\magiicat} sample, the drop in {\MgII} equivalent width at $D/R_{\rm vir}\gtrsim0.3$ is more pronounced in lower mass halos ($M_{\rm vir}<10^{12}~M_{\odot}$) and shallower in higher mass systems, implying inner CGM self-similarity across mass and an outer halo cool gas depletion that scales with halo mass. Project AMIGA observations of M31 further show that within ${\sim0.1R_{\rm vir}}$ (corresponding to 30~kpc) low and intermediate ionisation column densities remain high and co-rotating, while beyond ${\approx0.25R_{\rm vir}}$ low ionisation column densities fall steeply while {\OVI} declines more gradually \citep{lehner20,lehner25}. This is similar to our results showing strong low ionisation co-rotation to $D/R_{\rm vir}\approx0.2$ and an {\OVI} decoupling beyond.  

Theory predicts these $D/R_{\rm vir}$ scales naturally. Inflows shock to the virial temperature once a stable virial shock forms, expected for $M_{\rm vir}\gtrsim10^{11.5}~M_{\odot}$, leading to suppressed cold-mode accretion beyond $\sim0.2-0.3R_{\rm vir}$ as cooling times lengthen \citep{birnboim03,dekel06}. Filamentary cold streams dominate the inner halo but diminish in mass flux with both increasing radius and halo mass \citep{keres05,vandevoort11}.  FIRE-2 zoom-in simulations pin down the cooling 
flow radius at $r_{\rm cool}\approx0.1R_{\rm vir}$, where hot ($T\sim10^{5.5}$K) inflow rapidly cools and becomes rotationally supported \citep{hafen22}.  \citet{trapp22} showed that, within this cooling radius, the co-rotational support dominates ($\sim100${\kms}), whereas radial transport is negligible ($\sim1-3${\kms}).  Beyond $0.2R_{\rm vir}$ the hot, pressure-supported CGM fails to co-rotate fully \citep{hafen22,trapp22}. Similarly, \citet{stern21} show that the CGM first virialises outside‐in, achieving hot, quasi-static conditions from $\sim0.5R_{\rm vir}$ inward to $\sim0.1R_{\rm vir}$, with the inner CGM only becoming long-lived and pressure-supported once $M_{\rm vir}\gtrsim10^{12}~M_{\odot}$.  

In COS–EDGES, our low and high $D/R_{\rm vir}$ samples have identical mean halo masses log$(M_{\rm h}/M_{\odot})_{\rm low}=11.58\pm0.19$ and log$(M_{\rm h}/M_{\odot})_{\rm high}=11.55\pm0.24$, confirming that the observed drop in co-rotation fraction and $v_{\rm abs}$ at $D/R_{\rm vir}\gtrsim0.2$ reflects a radial dependence and not a mass dependence.  Anchoring our kinematic boundary at \(D/R_{\rm vir}\sim0.2\) to the self-similar {\magiicat} results and AMIGA's inner-CGM mapping yields a unified picture: the inner halo is fed by high angular momentum cold flows and recycled fountains, while beyond $\sim 0.2-0.3R_{\rm vir}$, diminished cold supply and the rise of a hot, pressure-supported phase produce the weakened rotation signatures, especially in {\OVI}.  

Our results further contribute to low ionisation {\MgII} kinematic studies that show the CGM commonly co-rotates along the major axes of galaxies, even extending to angles well beyond the major axis plane, out to impact parameters of  $\sim100$~kpc \citep{steidel02,kacprzak10,kacprzak11kin,bouche13,bouche16,diamond-stanic16,ho17, rahmani18, martin19,lopez20,nateghi24GFII}. 
\citet{nateghi24GFII} further examined the multi-phase CGM around 27 galaxies and found that low-ionisation CGM is more kinematically consistent with the galaxy rotation than high-ionisation gas and showed a monotonic decline in the fraction of co-rotating gas with increasing ionisation potential. However, these results were based on a range of galaxy properties and galaxy-quasar geometries. Here, for our more restrictive edge-on major axis sample, we also find a drop of $\sim16-26$\% in equivalent width co-rotation fraction between {\MgII} and {\OVI}, consistent with those reported by \citet{nateghi24GFII}.

We find that the velocity widths corresponding to 50\% of the total optical depth ($\Delta v_{50}$) increase with increasing ionisation potential at high $D/\Rv$. The values also increase as the gas is found closer to the galaxy, with the exception of {\OVI}, which has a constant $\Delta v_{50}$ (and $\Delta v_{90}$) regardless of distance. This is consistent with \citet{nielsen17} who showed that velocity widths of {\MgII} absorption are narrower and dependant on properties such as azimuthal angle and galaxy colour, compared to {\OVI} which exhibits similar velocity widths with no dependencies. These results suggest that {\OVI} likely probes an outlying, dynamically hot component of the halo that does not participate in the disk’s rotation at large radii and traces the galaxy mass better, since the {\OVI} ion fraction peaks near $T \approx 10^{5.5}$~K, close to the virial temperature of $M_{\rm H} \sim 10^{12}~M_{\odot}$ systems, making it a sensitive tracer of the halo potential \citep{oppenheimer16,ng19,dutta25}.  This was further demonstrated in simulations, where {\OVI} was found to trace both accreting/co-rotating gas as well as the diffuse component within the halo \citep{kacprzak19kine} or even residing in the IGM \citep{ho21,bromberg24}. 

\citet{werk16} showed that {\OVI} absorbers separate into two distinct kinematic families. The first, the narrow/low ionisation–aligned class, contains {\OVI} absorption that has Doppler parameters $b \lesssim 25$~{\kms} and centroid offsets $\Delta v \lesssim 20$~{\kms} from the low-ion absorption. The second, the broad/no low ionisation class, contains {\OVI} components with no accompanying low ions, with $b \gtrsim 40$~{\kms} and velocity offsets up to $\pm100$ {\kms}. \citeauthor{werk16} interpret the former as multi-phase material moving coherently with the cool gas as either planar inflow or recycled fountain flow, whereas the latter traces hotter, turbulent coronal gas ($T \approx 10^{5.5}$~K) that is kinematically decoupled from the disk. Our COS–EDGES sight-lines at $D/R_{\rm vir} \le 0.2$, where {\OVI} is narrow and more closely velocity-matched to {\MgII}/{\CII}, match well with the narrow/low–aligned population. In contrast, the broader, more symmetric {\OVI} wings that dominate beyond $D/R_{\rm vir} > 0.2$ are consistent with their broad/no–low population. 

Our results agree with earlier work by \citet{kacprzak19kine}, who found that {\OVI} detected along galaxy major axes shows little correlation with disk rotation; indeed, most of their sight-lines were at larger radii ($D/R_{\rm vir}\gtrsim0.2$), where we see a weaker kinematic coupling to the galaxy. More recently, \citet{ho25} demonstrated that while individual {\OVI} components need not trace the exact disk rotation curve, the bulk of the {\OVI} gas almost never counter-rotates. In their sample, {\OVI} tied to low-ion gas ({\SiII}/{\SiIII}), typically probing the inner CGM, was almost exclusively co-rotating, whereas {\OVI} without low-ion counterparts (typically tracing more distant gas) shows a much higher incidence of counter-rotation and random motions. Up to $\sim80\%$ of low-ion absorbers lack sufficient angular momentum for a stable orbit, compared to only $\sim45\%$ of {\OVI} absorbers. Together, these results extend and reinforce our picture of a co-rotating, inflowing inner CGM and a more turbulent, kinematically decoupled outer halo. In addition, detailed cloud-by-cloud analyses indicate that broad {\OVI} at large radii arises in discrete, collisionally ionised structures embedded within a hotter, more tenuous medium; in lower-mass halos, this ambient medium may be too hot or diffuse to produce strong {\OVI} by itself \citep{sameer24}. This is supported by simulations that predict that for $M_{\rm halo}\lesssim10^{12}{\rm M}_\odot$ systems, {\OVI} is produced in localised warm-hot clouds or interface layers, while the volume-filling phase is hotter ($T\gtrsim10^6$~K) and traced by higher ions or X-rays \citep{oppenheimer16,nelson18,bradley22}. 

These previous works support our radius–dependent results where {\OVI} likely originates from two distinct environments: narrow, low-ion-aligned gas that co-rotates with the disk, and broad, low-ion–free gas that is hotter, kinematically decoupled, arises in discrete collisionally ionised structures embedded within a hotter halo. The fact that we observe {\OVI} co-rotation fraction of $\sim85$\%, with a optical depth-weighted median velocity matching the rotation speed of the galaxy, at low $D/{\Rv}$ suggests that near the galaxy, even this hot/warm medium can be partially entrained by the disk’s gravitational potential or by mixing with inflow. However, at larger radii, {\OVI} becomes increasingly decoupled and could arise in discrete, collisionally ionised substructures, warm volume-filling gas, and/or from the IGM.

Overall, the kinematic signatures detected in our sample, combined with the major axis geometry, favour a cold‐accretion scenario in the inner halo. Cosmological simulations consistently show that filamentary inflows carry high specific angular momentum into galaxy halos and settle into extended, rotating structures. In zoom‐in simulations, cold‐mode gas delivers three to four times the dark‐matter angular momentum and forms coherent cold‐flow disks tens of kiloparsecs across \citep{stewart11,stewart13,stewart17,oppenheimer18b}. The EAGLE simulations demonstrated that this cool inflow remains confined to the galaxy plane, co‐rotating with the disk out to $\sim60$kpc \citep{ho19}, and TNG50 simulations reproduce inspiralling streams around $M_{\rm h}=10^{10-11} M_\odot$ halos \citep{nelson20}. FIRE particle‐tracking further reveals that freshly accreted IGM gas enters the outer disk already in rotation, confirming that pristine filaments directly feed galactic disks \citep{hafen19,hafen22,trapp22}. Cosmological zoom-in simulations further indicate that inflowing gas approaches Milky Way–mass disks with angular momentum comparable to that at the disk edge and low radial velocities, settling into extended, co-rotating structures tens of kiloparsecs across before slowly transporting inward at $\sim1-3$\,km\,s$^{-1}$ to fuel star formation \citep{trapp22}.  Such extended cold-flow disks naturally produce the strong co-rotation signatures at $D/R_{\rm vir}\lesssim0.2$ seen in COS–EDGES, where optical-depth–weighted velocities match disk rotation.  The slow radial inflow implies that beyond $D/R_{\rm vir}>0.2$, gas retains only partial kinematic coupling, consistent with the observed drop in $v_{\rm abs}$ and co-rotation fraction.  In addition, feedback-driven torques and spiral-arm oscillations in simulations generate multi-component, turbulent profiles in the inner halo \citep{trapp22}, matching the broader $\Delta v_{50}$ widths we observe close to galaxies. While we do detect these kinematic signatures, galactic fountains enrich and recycle CGM gas in a similarly co‐rotating fashion \citep[e.g.,][]{marinacci10,marasco2012,fraternali17,armillotta16,grand19}. 

In FIRE and Auriga simulations, star‐formation–driven outflows launched along the minor axis cool and fall back toward the disk, spinning up as they reaccrete and forming metal‐rich, planar inflows \citep{angles17,muratov17,marinacci19}. Idealized models of hot‐mode accretion show that mixing between fountain ejecta and ambient corona can transfer angular momentum back into the returning gas \citep{stern24}, so that even recycled material rejoins the disk with significant rotation. These fountain flows complement the cold filaments by sustaining a multi‐phase, co‐rotating inner halo. If some of the detected low-ion gas originates from past outflows that have cooled and re-accreted, it would likely be metal-enriched \citep{peroux20,weng24}. Simulations also predict that co-rotating inflow near the disk should exhibit lower metallicities compared to recycled fountain gas \citep{hafen19,hafen22,trapp22}, suggesting future metallicity measurements of co-rotating CGM absorbers could distinguish between pristine and recycled channels feeding galaxies.
Interestingly, recent work by \citet{nateghi24GFII} identified major axis co-rotating low-ion gas with lower metallicity, which may trace more pristine inflow along cosmic filaments, whereas co-rotating gas with higher metallicity may signify recycled, co-planar accretion of prior outflows. Thus, it is possible that both scenarios may be occurring in our sample and our future works will address these key questions by quantifying the cloud-by-cloud properties \citep[e.g.,][]{sameer24}, including metallicities and by performing detailed kinematic modelling.  

\section{Conclusions}
\label{sec:conclusion}
We introduce the COS-EDGES survey, which presents a new sample of absorption line systems observed with COS/{\it HST} and VLT/UVES to address how multi-phase gas flows toward galaxies from the CGM. Our selected sample consists of $z\sim0.2$, isolated, edge-on, emission-line galaxies with a background quasar located along their major axis between $D=13-37$~kpc and $D/\Rv = 0.12-0.31$. In paper introducing this sample, we examine the kinematic connection between the galaxies and their multiphase CGM, with a focus on how the kinematics change with increasing $D/R_{\rm vir}$. Our results can be summarised as follows:

\begin{itemize}
    \item For individual systems, most of the absorption is consistent with the galaxy's direction of rotation. Low ionisation absorbers appear to more consistently align with disk rotation, but several galaxies (e.g., G1, G2, G4, G6, G8) show high-ion absorption that are misaligned or counter-rotating. Misalignment appears to increases with increasing distance from the host galaxies. 
    
    \item The multiphase CGM gas appears to be gravitationally bound. Low ionisation gas in eight of nine galaxies, and high ionisation absorption in six of nine galaxies, occurs below the escape velocities calculated at the virial radii. For all ions, the CGM is within the escape velocities calculated at the impact parameter for each system, showing that all of the CGM appears to be retained by the halo potential.

    \item At low  $D/R_{\rm vir}$ ($\leq 0.2 R_{\rm vir}$, $0.12\leq$$D/R_{\rm vir}$$\leq0.20$) the stacked profiles {\MgI} and {\MgII} exhibits single-sided, rotation-matched profiles, while {\HI}, {\CII}, {\CIII} and {\OVI} have a small amount of absorption in the opposite direction of rotation. 
    
    \item At high $D/R_{\rm vir}$ ($>0.2 R_{\rm vir}$,  
 $0.21\leq$$D/R_{\rm vir}$$\leq0.31$) the stacked profiles broaden and become more symmetric about systemic velocity, but {\MgII} and {\MgI} still show a dominant co-rotating component. In contrast, {\OVI} and {\HI} exhibit substantial absorption on both sides of the galaxy systemic velocity, indicating a weaker rotational connection to the disk.

    \item The optical depth weighted median velocity, $v_{\rm abs}$, is $\sim$110 {\kms} for all ions within the inner halo, but drops to 40~{\kms} at larger $D/R_{\rm vir}$, with the largest drop for {\OVI} at $\sim 20$~{\kms}. When normalised to the maximum rotation of the galaxy, all ions are consistent with that velocity at low $D/R_{\rm vir}$, but decrease to about 60\% of the maximum rotation speed for the low ionisation gas and 20\% for {\OVI}. 

   \item The 50\%-optical-depth width ($\Delta v_{50}$) for low ions (e.g., {\MgII}, {\CII}) is up to $\sim1.8$ times larger in the inner halo compared to the outer, indicating a more turbulent or multi-component medium near the galaxy. The $\Delta v_{50}$ does not change with distance for the high ions ({\CIII} and {\OVI}). The 90\%-optical-depth width ($\Delta v_{90}$) shows a modest decline with radius for low ions while  {\CIII} and {\OVI} remain unchanged. At high $D/\Rv$, $\Delta v_{50}$ and $\Delta v_{90}$ increases with increasing ionisation potential.

    \item The equivalent width co-rotation fraction declines systematically with both ionisation potential and radius: from unity to 0.93 for {\MgII} from the inner to outer halo and from 0.84 to to 0.63 for {\OVI}. We find that the slope, \dfcorot, becomes steeper at large $D/R_{\rm vir}$, which further supports the picture of a radially-stratified, multi-phase CGM.
\end{itemize}

These results are suggestive of a picture in which dynamically broad co-rotating multi-phase gas dominates the inner CGM, possibly tracing extended inflow or recycled accretion streams, whereas the outer CGM is characterised by a lower relative velocity dispersion component that is more weakly aligned with the rotation of the galaxy. The warm phase ({\OVI}) lags the most and could be composed of co-rotating, lagging and volume-filling components. This program sheds light on these radial differences and the complex pathways by which galaxies exchange mass and angular momentum with their CGM. Future work will examine a comparison between the galaxy ISM and CGM metallicities, gas flow models using higher resolution imaging and CGM mapping in emission to further address how galaxies acquire their gas.

\paragraph{Acknowledgments}
This work is dedicated to the memory of Jacqueline Bergeron, whose pioneering research on the galaxy-CGM absorption connection inspired and paved the way for a multitude of studies such as ours. Some of the data presented herein were obtained at the W. M. Keck Observatory, which is operated as a scientific partnership among the California Institute of Technology, the University of California, and the National Aeronautics and Space Administration. The Observatory was made possible by the generous financial support of the W. M. Keck Foundation. Observations were supported by Swinburne Keck program 2022B\_W200. The authors wish to recognize and acknowledge the very significant cultural role and reverence that the summit of Maunakea has always had within the indigenous Hawaiian community. We are most fortunate to have the opportunity to conduct observations from this mountain.
Some of the data presented herein were based on observations collected at the European Organisation for Astronomical Research in the Southern Hemisphere under ESO programmes 105.20FN.001, 105.20FN.002 and 108.22F4.002. This paper includes data gathered with the 6.5 meter Magellan Telescopes located at Las Campanas Observatory, Chile.

\paragraph{Funding Statement}
This work was supported by Space Telescope Science Institute grant HST-GO-17541, which funded BDO and CWC. 



\paragraph{Data Availability Statement}
The data underlying this paper will be shared following mutually agreeable arrangements with the corresponding authors.



\printendnotes



\bibliography{example}

\appendix
\section{Individual Galaxy Gas–CGM Kinematic Comparisons}

In this section, we show the remaining galaxies not shown in the manuscript (G1, G2, G3, G4, G5, G6 and G9). 

For each figure, the DECaLS $grz$ colour image of galaxy (left) and quasar (right) are shown.  The coloured circle data points show the galaxy rotation curves derived from nebular emission lines. Galaxies G2 and G5 exhibit shallow rotation curves likely due to their low star-formation rates ($SFR=0.10$ and $0.39~M_\odot$~yr$^{-1}$, respectively) producing faint emission lines, limiting the number of spatial apertures with sufficient signal-to-noise to derive reliable velocity centroids. The y-axes show both the line-of-sight velocity in {\kms} (right-side) as well as the peak rotation-normalised velocity (left side). The x-axes show the projected distance from the galaxy towards the quasar in kpc ($D$; lower axis) and in virial radius-normalised distance ($D/R_{\rm vir}$; top axis). The direction towards the quasar is in the positive x-axis direction and the direction of galaxy rotation towards the quasar is in the positive y-axis direction above the horizontal dashed line.  The vertical dotted line shows the projected distance of the quasar, which is also labelled. At the projected distance of the quasar, the spectra of individual CGM absorption lines are shown horizontally in order of increasing ionisation potential and are offset for clarity. Not all of the absorption lines detected in the quasar spectra are shown, but we have chosen to display a subset that are common for each galaxy ({\MgI}, {\HI}, {\OI}, {\MgII}, {\CII}, {\NII}, {\CIII} and {\OVI}). See figure captions for a discussion of each system.

\begin{figure*}
	\includegraphics[width=\textwidth]{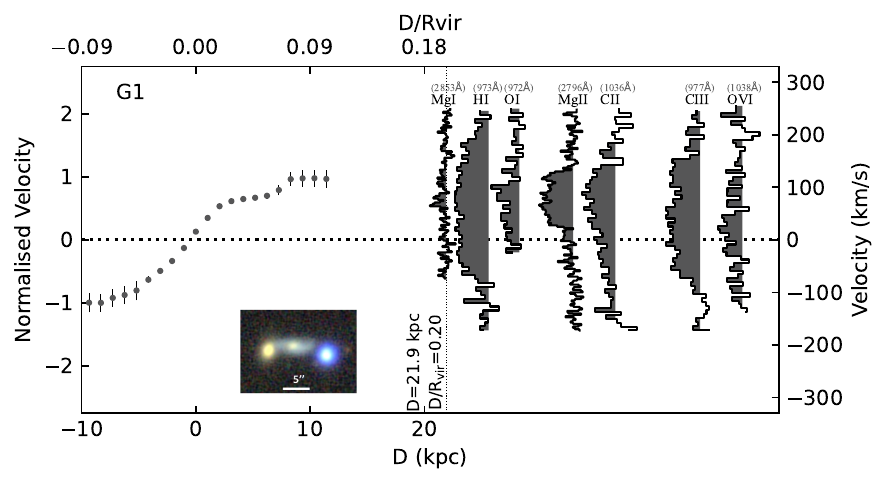}
    \caption{Same as Figure~\ref{fig:G7kine} but for G1, having a $D=21.9$~kpc and $D/\Rv = 0.20$ with a maximum rotation speed of $\sim110$~{\kms} in the direction of quasar. The low ionisation absorption ({\MgI}, {\OI}, {\MgII}) is consistent with the galaxy’s rotation direction, with absorption velocities ranging between the galaxy systemic velocity and the maximum rotation speed. {\HI}, {\CII}, {\CIII} and {\OVI} have absorption at the maximum rotation speed of the galaxy, with a small fraction of their absorption in the opposite direction of rotation.    
    }
    \label{fig:G1kine}
\end{figure*}
\begin{figure*}

    \includegraphics[width=\textwidth]{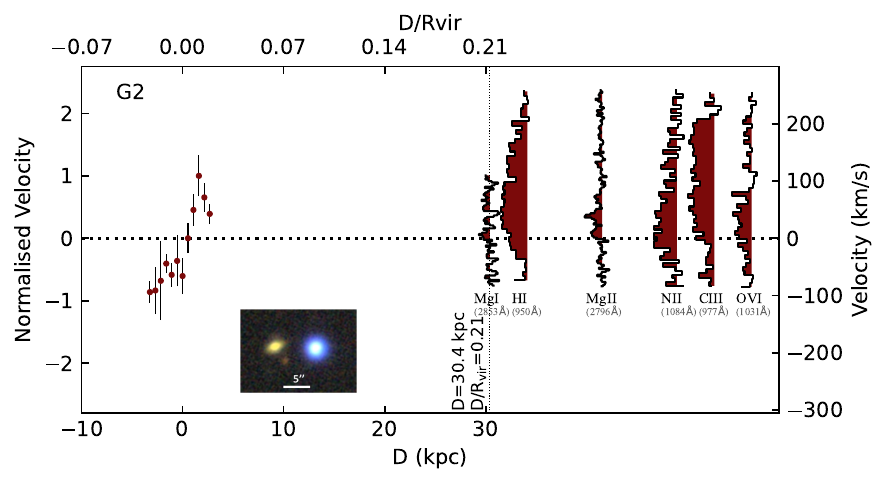}
    \caption{Same as Figure~\ref{fig:G7kine} but for G2, having a $D=30.4$~kpc and $D/\Rv = 0.21$, with a maximum rotation speed of $\sim100$~{\kms} in the direction of quasar. The low ionisation absorption ({\MgI}, and {\MgII}) is consistent with the galaxy’s rotation direction but at velocities lower than the maximum rotation speed of the galaxy. {\HI}, {\NII}, {\CIII} and {\OVI} have a small fraction of their absorption in the opposite direction of rotation.}
    \label{fig:G2kine}
\end{figure*}
\begin{figure*}
	\includegraphics[width=\textwidth]{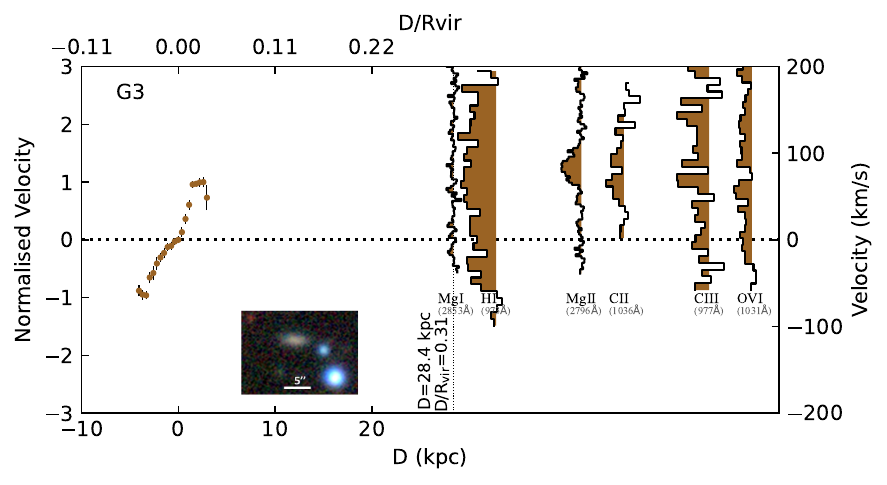}
    \caption{Same as Figure~\ref{fig:G7kine} but for G3, having a $D=28.4$~kpc and $D/\Rv = 0.31$, with a maximum rotation speed of $\sim65$~{\kms} in the direction of quasar. The low ionisation absorption ({\MgI}, {\MgII}, {\CII}) is consistent with the galaxy’s rotation direction and maximum rotation speed. {\HI}, {\CIII} and {\OVI} have a small fraction of their absorption in the opposite direction of rotation.}
    \label{fig:G3kine}
\end{figure*}
\begin{figure*}
	\includegraphics[width=\textwidth]{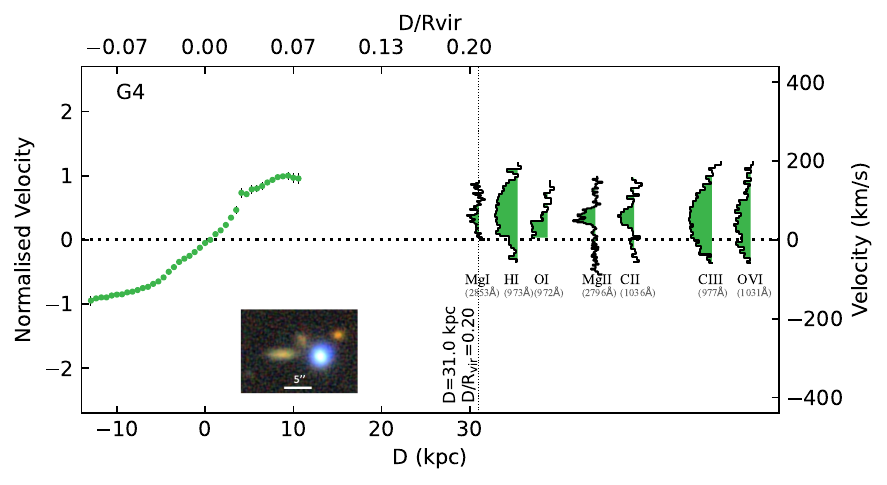}
    \caption{Same as Figure~\ref{fig:G7kine} but for G4, having a $D=31.0$~kpc and $D/\Rv = 0.20$, with a maximum rotation speed of $\sim155$~{\kms} in the direction of quasar. The low ionisation absorption ({\MgI}, {\OI}, {\MgII}, {\CIII}) is consistent with the galaxy’s rotation direction and slower than the maximum galaxy rotation speed. {\HI}, {\CIII} and {\OVI} have a small fraction of their absorption in the opposite direction of rotation.}
    \label{fig:G4kine}
\end{figure*}

\begin{figure*}
	\includegraphics[width=\textwidth]{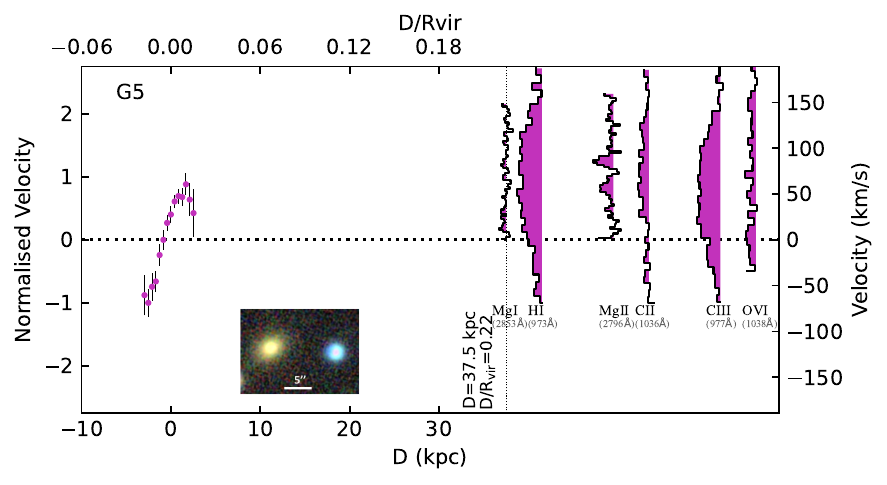}
    \caption{Same as Figure~\ref{fig:G7kine} but for G5, having a $D=37.5$~kpc and $D/\Rv = 0.22$, with a maximum rotation speed of $\sim60$~{\kms} in the direction of quasar. The low ionisation absorption ({\MgI}, {\MgII}, {\CIII}) is consistent with the galaxy’s rotation direction, with velocities above and below the maximum galaxy rotation speed. {\HI}, {\CIII} and {\OVI} have a small fraction of their absorption in the opposite direction of rotation.}
    \label{fig:G5kine}
\end{figure*}
\begin{figure*}
	\includegraphics[width=\textwidth]{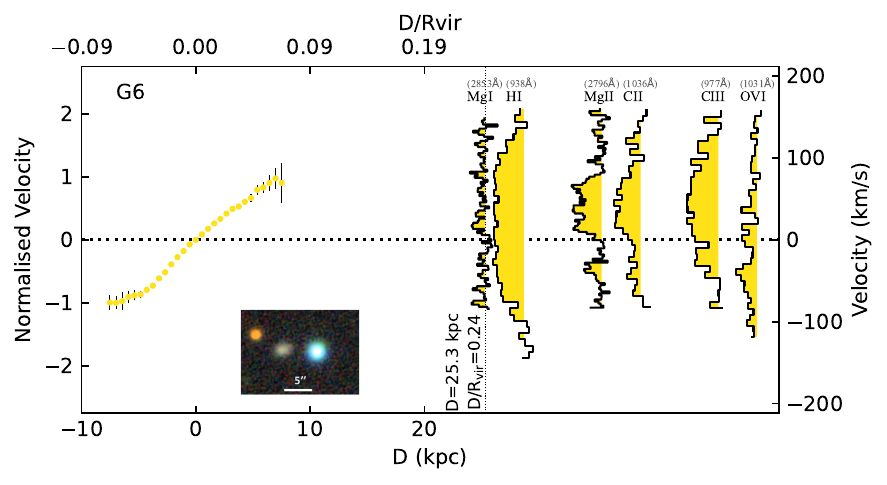}
    \caption{Same as Figure~\ref{fig:G7kine} but for G6, having a $D=25.3$~kpc and $D/\Rv = 0.24$, with a maximum rotation speed of $\sim75$~{\kms} in the direction of quasar. The bulk of the low ionisation absorption ({\MgI}, {\MgII}, {\CIII}) is consistent with the galaxy’s rotation direction, with velocities at and below the maximum galaxy rotation speed. Some low ionisation absorption is also found in the opposite direction of rotation. {\HI}, {\CIII} and {\OVI} have a small fraction of their absorption in the opposite direction of rotation.}
    \label{fig:G6kine}
\end{figure*}

\begin{figure*}
	\includegraphics[width=\textwidth]{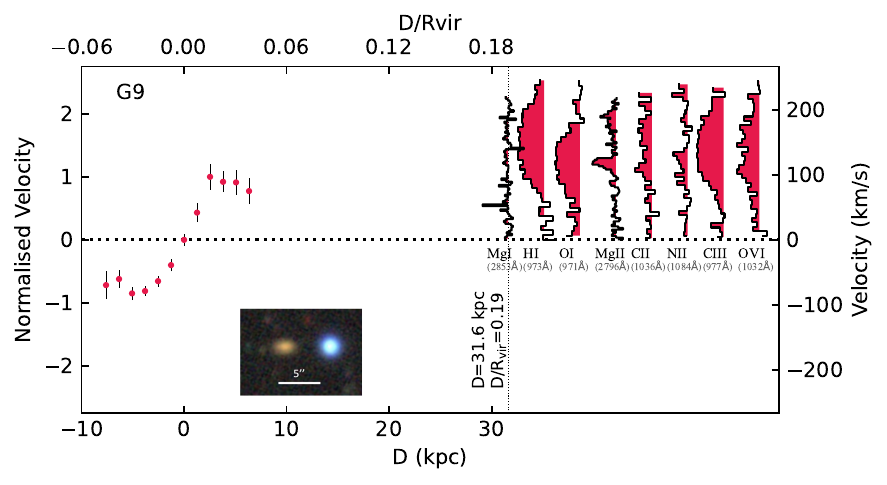}
    \caption{Same as Figure~\ref{fig:G7kine} but for G9, having a $D=31.6$~kpc and $D/\Rv = 0.19$, with a maximum rotation speed of $\sim100$~{\kms} in the direction of quasar. All of the ions have velocities consistent with the galaxy’s rotation direction. }
    \label{fig:G9kine}
\end{figure*}

\end{document}